\documentclass[authoryear,12pt]{elsarticle}
\usepackage{lipsum}
\usepackage{color}

\newcommand{\mprime}{m'_{\alpha\beta}}

\usepackage{graphicx}
\usepackage{amsmath}
\usepackage{amssymb}
\usepackage{verbatim}
\usepackage{url}
\usepackage{gensymb}
\usepackage{booktabs}
\usepackage{setspace}
\usepackage{tabularx}
\usepackage{tabulary}
\usepackage{hhline}
\usepackage{multirow}
\usepackage{multicol}
\usepackage{float}
\usepackage{bm}
\usepackage{amsmath}
\usepackage[subnum]{cases}
\usepackage{amssymb}
\usepackage{subfigure}
\usepackage{array}
\usepackage{caption}
\newcolumntype{P}[1]{>{\centering\arraybackslash}p{#1}}
\journal{European Journal of Mechanics/A Solids, accepted for publication}
\begin{document}
\begin{frontmatter}
 \title{On the effect of slip transfer at grain boundaries on the strength of FCC polycrystals}
 \author{E. Nieto-Valeiras$^{1,2}$}
 \author{S. Haouala$^{1}$}
 \author{J. LLorca$^{{1,2,}^{*}}$}

 \address{$^1$IMDEA Materials Institute, C/ Eric Kandel 2, 28906 Getafe, Madrid, Spain}
 \address{$^2$Department of Materials Science, Polytechnic University of Madrid/Universidad Polit\'ecnica de Madrid, E. T. S. de Ingenieros de Caminos, 28040 Madrid, Spain}
 \cortext[cor1]{Corresponding author; Email address: javier.llorca@imdea.org }
\begin{abstract}
The effect of slip transfer on the flow strength of various FCC polycrystals was analyzed by means of computational homogenization of a representative volume element of the microstructure. The  crystal behavior was governed by a physically-based crystal plasticity model in the framework of finite strains where slip transfer at grain boundaries was allowed between slip systems suitably oriented according to geometrical criteria. Conversely, slip transfer was blocked if the conditions for slip transfer were not fulfilled, leading to the formation of dislocation pile-ups. All the model parameters for each material were identified from either dislocation dynamics simulations or experimental data from the literature. Slip transfer  led to a reduction in the flow stress of the polycrystals (as compared with the simulations with opaque grain boundaries) which was dependent on the fraction of translucent and transparent grain boundaries in the microstructure. Moreover, dislocation densities and Von Mises stresses were much higher around opaque grain boundaries, which become suitable places for damage nucleation. Finally, predictions of the Hall-Petch effect in Al, Ni, Cu and Ag polycrystals including slip transfer were in better agreement with the literature results, as compared with predictions assuming that all grain boundaries are opaque, particularly for small grain sizes ($<$ 20 $\mu$m).
\end{abstract}
\end{frontmatter}
%
\section{Introduction}
\label{sec_intro}

Metallic materials stand for ideal candidates for a wide range of structural applications in automotive, aerospace, chemical, construction and biomedical sectors owing to their high stiffness, formability and ductility. However, the strength of pure metals is very low for many applications due to the development of plastic deformation by dislocation slip. Hence, they have to be strengthened by obstacles that hinder dislocation motion and grain boundaries (GBs) stand among the strongest barriers to dislocation slip in polycrystals.  Following \cite{Bayerschen2016a}, the interactions of dislocations with GBs can be grouped in three categories. Opaque or impenetrable GBs do not allow the propagation of dislocation slip through the boundary, leading to the formation of dislocation pile-ups and local stress concentrations \citep{friedman1998continuum,Bieler2014}. Moreover, the accumulation of dislocations at the GB induces strong strain gradients in order to preserve the displacement continuity between neighbor grains \citep{hughes2003geometrically}. Conversely, transparent GBs are found when all the active slip systems are suitably aligned and the dislocations gliding in the incoming  grain can be  transmitted to the neighbor grain with no induced stress concentration. Finally,  dislocations in some active slip systems can easily propagate through the boundary while others are blocked  in the case of translucent GBs. This phenomenon leads to stress concentrations at the GB that depend on the number and intensity of slip in the systems blocked at the GB.

The effect of GBs on the strength of polycrystals was recognized in the 1950's, leading to the phenomenological Hall-Petch law \citep{hall1951deformation,petch1953cleavage} that relates the yield strength of the polycrystal, $\sigma_y$ with the average grain size, $\bar{D}_g$, according to \begin{equation}
\sigma_y = \sigma_{\infty} + C_{HP} \, {\bar{D}_g}^{-x}
\label{eq_hallpetch}
\end{equation}

\noindent where $\sigma_{\infty}$ is the yield stress of a polycrystal with a very large grain size, $C_{HP}$ a material constant and $x$ a scaling exponent in the range -0.5 to -1  for many metallic alloys \citep{raj1986compilation,dunstan2013scaling,dunstan2014grain,li2016hall}. This expression is supported by different theoretical models. For instance, \cite{Ashby1970} proposed a physically-based model based on the strain incompatibility between  grains with different orientations within a polycrystal. In this model, the total density of dislocations arises from two distinct contributions termed statistically stored dislocations (SSDs) and geometrically necessary dislocations (GNDs). The contribution of GNDs depends on the grain size and concentrates at the GBs whereas the density of SSDs is  size independent. Other theoretical approaches also  provided a rational explanation of the Hall-Petch law with a grain representation formed by a soft core surrounded by hard shell around the grain boundary \citep{kocks1970relation,Hirth1972}.

Understanding the mechanical behavior of polycrystals has progressed rapidly in recent years through the combination of computational homogenization and crystal plasticity constitutive models \citep{Segurado2018}. Within this framework,  two different strategies have been used to study the effect of GBs on the strength of polycrystals based on either strain-gradient  or physically-based crystal plasticity models. In the former, a length scale is introduced in the crystal plasticity constitutive equation through the plastic strain gradients, which are related to the density of GNDs near the grain boundaries through the Nye tensor \citep{Nye1953}. Different investigations have used this approach to analyze the effect of grain size on the strength of polycrystals \citep{AB00, CBA05, BBG07, BER10, LN16}. Nevertheless, the comparisons with the abundant available experimental data were not fully convincing because  most of the simulations were limited to small representative volume elements (RVEs) of the microstructure having only a few dozens of grains due to computational limitations. They were overcome by  \cite{Haouala2020}, who determined the effect of grain size on the strength of different FCC polycrystals by means of computational homogenization  of RVEs containing several hundred grains using a Fast Fourier Transform in combination with a strain gradient crystal plasticity model. The density of GNDs resulting from the incompatibility of plastic deformation among different crystals was obtained from the Nye tensor, which was efficiently obtained from the curl operation in the Fourier space. The simulation results were in good agreement with the experimental data  for Cu, Al, Ag and Ni polycrystals for grain sizes $>$ 20 $\mu$m. Similar results were also obtained by \cite{Rubio2019}  through computational homogenization of equivalent RVEs using the finite element method and a physically-based crystal plasticity model that takes into account the storage of dislocations near the grain boundaries, following the approach developed by \cite{Haouala2018}. Interestingly, the predictions of the physically-based model also overestimated the strength of the polycrystals when the average grain size was $<$ 20 $\mu$m. It was argued that the differences between experimental data and simulation results (using either strain gradient or physically-based models) arose because all GBs were assumed to be opaque in the simulations. 

\begin{figure}[h]
	\centering
	\includegraphics[width=0.8\textwidth]{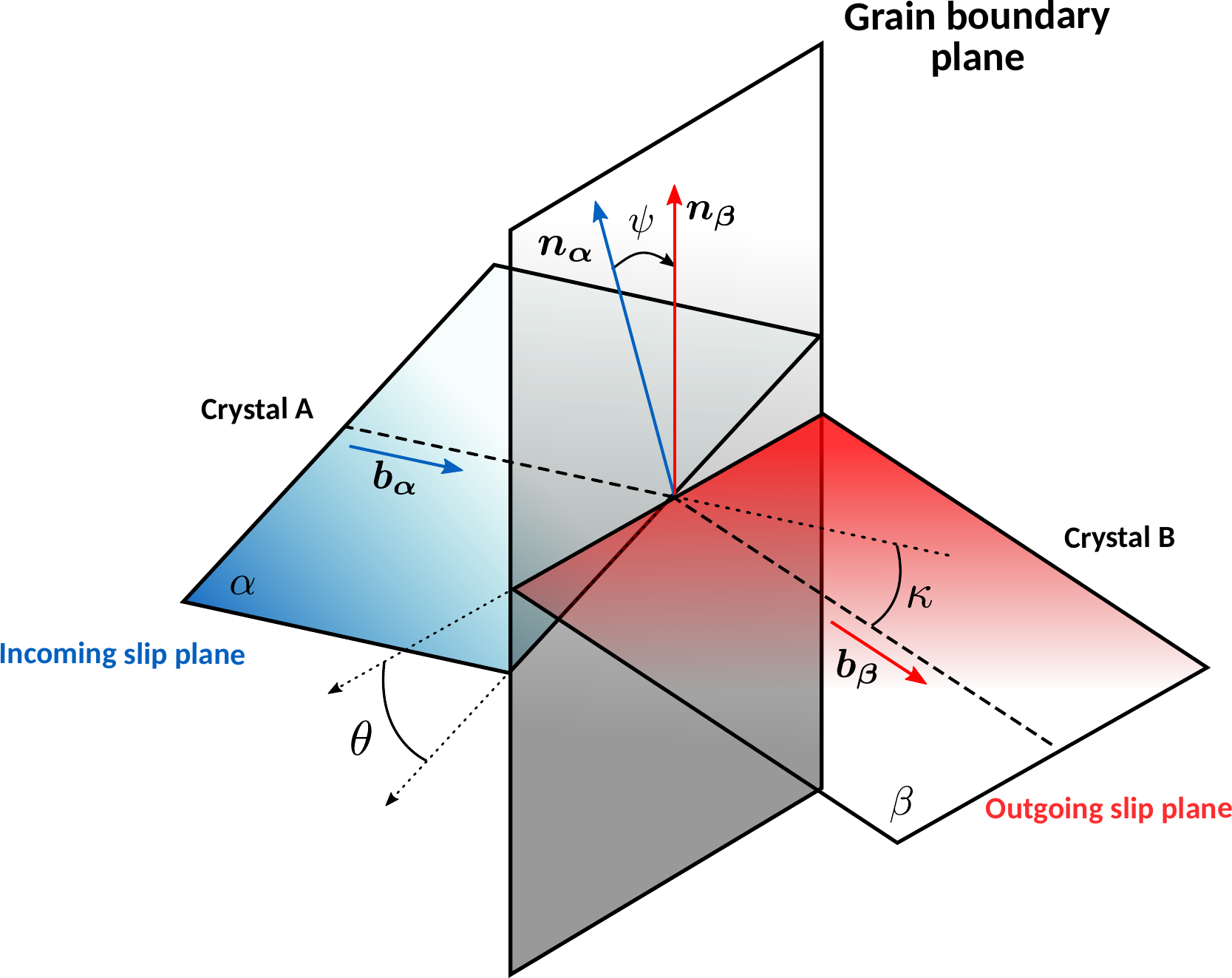}
	\caption{Geometrical parameters to assess slip transfer across the GB.}
	\label{st_geometry}
\end{figure}

There is ample experimental evidence indicating that  slip transfer through the GBs can take place \citep{Shen1988,Lee1989} and is often associated with a good alignment between incoming and outgoing slip systems. Thus, geometrical criteria based on the orientation of the grains and slip systems across the boundary have been used to analyze slip transfer. The geometrical parameters used to predict slip transfer between two slip systems on either side of a GB are defined in Fig. \ref{st_geometry}. They are the angle $\kappa$ between the Burgers vectors of the dislocations in the incoming ($\bm{{b}_{\alpha}}$) and outgoing ($\bm{{b}_{\beta}}$) slip plane, the angle $\psi$ between the slip plane normals ($\bm{{n}_{\alpha}}$ and $\bm{{n}_{\beta}}$)  and the angle $\theta$ between the two slip plane intersections with the GB plane.

The experimental evidence indicates that slip transfer takes place more easily in low-angle GBs, i.e. when the misorientation angle between adjacent grains is below $\approx$ 15$^\circ$ \citep{Read1950a}. Nevertheless, slip transfer is not guaranteed at low-angle GBs, not it is  impossible at large GB misorientations. Based on {\it in situ} TEM deformation experiments, \cite{Lee1989} proposed another slip transfer criterion from grain A to grain B. The so-called $LRB$ parameter was expressed by

\begin{equation}\label{LRB}
	LRB = \cos\theta  \cos\kappa
\end{equation}

\noindent and slip transfer was likely when $LRB$ was close to 1, indicating good alignment between the slip planes at the GB and a small magnitude of the residual Burgers vector left at the grain boundary, $ \Delta b_{\alpha\beta} = |\bm{b}_{\alpha}-\bm{b}_{\beta}|$. 

Another criterion for slip transmission was proposed by \cite{Luster1995}, who  emphasized the importance of the good alignment between the incoming and outgoing slip systems as well as a small magnitude of the residual Burgers vector left at the grain boundary through the geometric compatibility factor $m'_{\alpha\beta}$,  defined as 

\begin{equation}
m'_{\alpha\beta}=\cos \psi \cos \kappa.
\label{mprime_criteria}
\end{equation}

Slip transfer was likely when $m'_{\alpha\beta}$ was close to 1, as it indicates full compatibility between the slip systems $\alpha$ and $\beta$ across the GB because both the slip directions and the slip planes are parallel and dislocations should be easily transmitted across the GB. On the contrary, the slip systems are incompatible if $m'_{\alpha\beta}=0$   because either the slip directions or slip planes are orthogonal.  This criterion has been  used to assess slip transfer and blocking across GBs in various investigations. For instance,  \cite{Hemery2018} studied slip transfer in  Ti-6Al-4V alloy polycrystals and found that high values of $m'_{\alpha\beta}$, combined with high resolved shear stress in the outgoing slip system, was a good indicator of the probability of slip transfer. \cite{Bieler2019}  and \cite{alizadeh2020criterion} analyzed slip transfer in over 250 grain boundaries in pure Al polycrystals and concluded that slip transfer is likely to occur when $m'_{\alpha\beta} >$ 0.9 and $ \Delta b_{\alpha\beta} < 0.35 |\bm{b}|$, or at values of $m'_{\alpha\beta}/\Delta b_{\alpha\beta}$ that exceed a threshold. Finally, \cite{Abuzaid2016} also concluded the small values of $\Delta b_{\alpha\beta} $ are also good indicators of slip transfer. \cite{ZBL20} analyzed these experimental data using artificial neural networks to conclude that optimum predictions of slip transfer (with an accuracy close to 90\%) can be achieved through the combination of two geometrical metrics (such as GB misorientation and $m'_{\alpha\beta}$).

The effect of these geometrical criteria on slip transfer was recently included in a physically-based crystal plasticity model that was used to study stress concentrations in either transparent, translucent or opaque GBs in bycrystals oriented for single and double slip \citep{Haouala2019}. The results of the numerical simulations were in good agreement with experimental observations of slip transfer and slip activation on Al GBs, indicating that the crystal plasticity model was able to capture the effect of GBs on the slip behavior of FCC polycrystals. This strategy is extended in this paper to determine the flow strength of different FCC polycrystals by means of computational homogenization of an RVE of the microstructure, leading to accurate predictions of the Hall-Petch effect in Al, Cu, Ni and Ag particularly for grain sizes $<$ 20 $\mu$m. Moreover, the influence of the GB character (either transparent, translucent or opaque) on the local stresses that may trigger damage is also highlighted.

\section{Physically-based crystal plasticity model}
\label{sec_cp_model}

The mechanical behavior of the single crystals in the polycrystal is governed by a physically-based, rate-dependent crystal plasticity model \citep{Haouala2018}, which was  modified  to include the effect of slip blocking or transfer between slip systems concurring at a GB \citep{Haouala2019}. The model was developed within the framework of finite deformations using the multiplicative decomposition of the deformation gradient, following standard crystal plasticity implementations \citep{Segurado2018}. 

The relationship between the  plastic shear deformation rate in the slip system $\alpha$, $\dot{\gamma}^{\alpha}$,  and the corresponding resolved shear stress, $\tau^{\alpha}$, is expressed by 

\begin{equation}
	\dot{\gamma^{\alpha}}=\dot{\gamma_0}\left(\dfrac{\left\lvert \tau^{\alpha}\right\rvert}{\tau_c^\alpha}\right)^{\frac{1}{m}} sgn(\tau^{\alpha})
\end{equation}

\noindent where $\dot{\gamma_0}$ is the reference shear strain rate, $m$ the strain rate sensitivity parameter and ${\tau_c}^\alpha$ is the critical resolved shear stress (CRSS) in the slip system $\alpha$. Standard values of $\dot{\gamma_0}$ and $m$ for FCC polycrystals can be found in Table \ref{tab_FCC_parameters} \citep{Haouala2018}.

The evolution of the CRSS during deformation follows a particularization of the Taylor model \citep{taylor1934mechanism} by  \cite{franciosi1980latent} which includes the contribution  between dislocations in different slip systems to the hardening according to

\begin{equation}
	\tau_c^\alpha=\mu b\sqrt{\sum_{\delta}q^{\alpha\delta}\rho^{\beta}}
	\label{eq_taylor_tau}
\end{equation}

\noindent where $\mu$, $b$ and $\rho^\delta$ stand for the shear modulus parallel to the slip plane, the Burgers vector and the dislocation density in slip system $\delta$, respectively. The dimensionless coefficients $q^{\alpha\delta}$ stand for  the different interactions between pairs of slip systems and were obtained by means of discrete dislocation dynamics simulations for FCC lattices where dislocation slip occurs in 12 \{111\}$<$110$>$ systems \citep{devincre2008dislocation,bertin2013hybrid}. Due to symmetry considerations, only six independent coefficients are necessary to determine the 12 $\times$ 12 coefficients of $q^{\alpha\delta}$ and they are depicted in Table \ref{tab_FCC_parameters}.

\begin{table} [!t]
	\caption {Parameters of the dislocation-based crystal plasticity model for FCC single crystals.}
	\begin{center}
			\begin{tabular}{p{8cm} c}
				\midrule	
				\multicolumn{2}{l}{\emph{Viscoplastic parameters}}\\ \midrule
				Reference shear strain rate $\dot{\gamma_{0}}$ ($s^{-1}$) & {$10^{-4}$}\\
				Strain rate sensitivity coefficient $m$ & 0.05 \\ \midrule
				\multicolumn{2}{l}{\emph{Dislocation interaction coefficients} ($q^{\alpha\delta}$)}\\ \midrule
				Self interaction & 0.122 \\
				Coplanar interaction & 0.122 \\				
				Collinear interaction & 0.657 \\
				Glissile junction & 0.137 \\
				Hirth lock & 0.084 \\
				Lomer-Cottrell lock & 0.118 \\											
				\midrule
				\label{tab_FCC_parameters}
			\end{tabular}
	\end{center}
\end{table}

The evolution of the dislocation density in each slip system, $\dot{\rho^{\alpha}}$, follows the Kocks-Mecking law \citep{kocks1975thermodynamics, kocks2003physics} that takes into account the balance between the generation and annihilation of dislocations. Nevertheless, the original idea of  Kocks-Mecking was modified by \cite{Haouala2019} to account for the formation of dislocation pile-ups in the slip system near a GB that does not allow for slip transfer leading to 

\begin{eqnarray}
\dot{\rho}^{\alpha}  &= & \frac{1}{b}\left(\frac{1}{\ell^{\alpha}}- 2 y_c \rho^{\alpha} \right)| \dot{\gamma}^{\alpha}|  \quad \mbox{if slip transfer is allowed for any} \ \beta \\
 \dot{\rho}^{\alpha}  & = & \frac{1}{b} \left(\max \left(\frac{1}{\ell^{\alpha}},\frac{K_{s}}{d_b}\right)- 2 y_c \rho^{\alpha} \right)| \dot{\gamma}^{\alpha}| \, \mbox{if slip transfer is blocked} \: \forall  \beta
\end{eqnarray}

\noindent where  $\beta$ stands for any slip system in the nearest neighbor grain. Eq. (6) stands for the standard Kock-Mecking law that includes two terms that control the generation and annihilation of dislocations. Dislocation generation in the slip system $\alpha$ depends on the dislocation Mean Free Path
(MFP), ${\ell^{\alpha}}$, which stands for the distance travelled by a dislocation segment before it is stopped by an obstacle and it is given by

\begin{equation}
	l^\alpha=\dfrac{K}{\sqrt{\sum_{\delta\neq\alpha}\rho^\delta}}
	\label{eq_MFP}
\end{equation}

\noindent where $\rho^\delta$ is the dislocation density in the slip system $\delta$ and $K$ is a dimensionless constant, known as the similitude coefficient, that relates the flow stress with the average wavelength of the characteristic dislocation pattern. It was estimated by \cite{sauzay2011scaling} and \cite{Rubio2019} for different FCC polycrystals. Dislocation annihilation is controlled by  current dislocation density in the system $\alpha$  and $y_c$, the effective annihilation distance between dislocations. This latter parameter was estimated for different FCC metals by \cite{Rubio2019} assuming equal densities of edge and screw dislocations. The annihilation distance for edge dislocations is very small  ($\approx 6b$) and independent of the metal while that for screw dislocations is a function of the stacking fault energy \citep{essmann1979annihilation,kubin2013dislocations}. Eq. (6) establishes the dislocation accumulation rate at any slip system in any Gauss point of the polycrystal that is either far away for any GB boundary or when -even if the Gauss point is close to a GB- slip transfer to another slip system $\beta$ is possible across the GB according to a geometrical criteria that has to be defined. 

In contrast, Eq. (7) provides the dislocation accumulation rate for a slip system at a Gauss point near a GB when slip transfer across the boundary is blocked according to the geometrical criteria. The critical distance at the GB that leads to an increase in the dislocation generation rate (and, thus, to the formation of a dislocation pile-up) is given by the condition $K_s/d_b > 1/\sqrt{\ell^{\alpha}}$ where $K_s$ is another dimensionless constant that determines the storage of dislocations at the GB and $d_b$ is the distance from the Gauss point to the nearest GB along the slip direction.   $K_s$ was also determined by means of 3D dislocation dynamics simulations in the presence of impenetrable grain boundaries in FCC polycrystals \citep{de2010grain} while $d_b$ depends on the location of the Gauss point and on the orientation of the crystal. The different coefficients of the crystal plasticity model for FCC Al, Ni, Cu and Ag as well as the elastic constants are depicted in Table  \ref{tab_single_crys}.
 
\begin{table} [t]
	\caption {Parameters of the dislocation-based crystal plasticity model  for the FCC crystals \citep{Haouala2018,Rubio2019}}
	\begin{center}
		\begin{tabular}{p{6cm} P{1.5cm} P{1.5cm} P{1.5cm} P{1.5cm}}
			\midrule
			& Cu & Al & Ni & Ag \\ \midrule
			\multicolumn{2}{l}{\emph{Elastic constants} (GPa)} \\ \midrule
			$C_{11}$ & 168.4 &  108 & 249 & 124\\
			$C_{12}$ & 121.4 & 61.3 & 155 & 93.7 \\
			$C_{44}$ & 75.4 & 28.5 & 114 & 46.1 \\
			Shear modulus $\mu$ & 30.5 & 25.0 & 58.4 & 19.5\\ \midrule
			\multicolumn{2}{l}{\emph{Dislocation parameters}}\\ \midrule
			Burgers vector $b$ (nm) & 0.256 & 0.286 & 0.250 & 0.288 \\
			Effective annihilation distance $y_c$ (nm) & 15 & 56 & 14 & 12.5\\
			Dislocation storage coefficient $K$ & 6 & 9 & 11 & 5\\
			Grain boundary storage coefficient $K_s$ & 5 & 5 & 5 & 5 \\ 						\midrule
			\label{tab_single_crys}
		\end{tabular}
			\end{center}
\end{table}

\section{Polycrystal homogenization framework}
\label{sec_ch_framework}
\begin{figure}[t]
	\centering
	\includegraphics[width=0.8\textwidth]{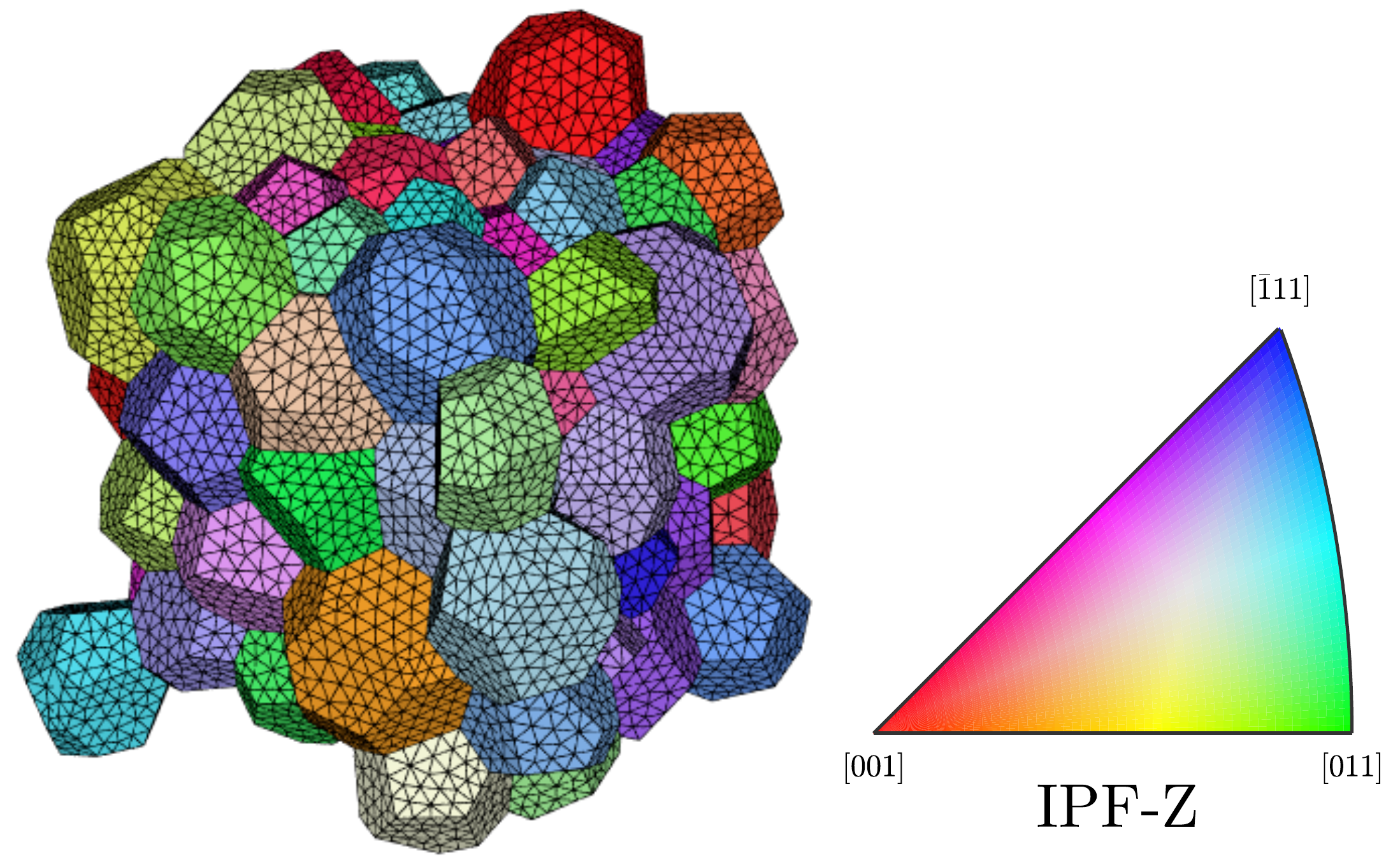}
	\caption{RVE of the microstructure including 100 grains discretized with 150000 second order tetrahedra. The colour of each grain corresponds to the orientation with respect to the Z axis as indicated in the inverse pole figure.}
	\label{RVE_mesh}
\end{figure}

The mechanical behavior of polycrystals with different grain sizes was determined by means of the finite element simulation of the deformation of cubic RVEs of the microstructure under uniaxial tension, following the standard procedures in full-field computational homogenization \citep{Segurado2018}. To this end, RVEs containing grains with random crystallographic orientations were generated using the open-source software Neper \citep{Quey2011a} and discretized with second order modified tetrahedra with \emph{Gmsh} \citep{Geuzaine2009} (C3D10M elements with 10 nodes in Abaqus). The shape of the grains was equiaxed and the grain sizes followed a lognormal distribution with average grain size $\bar{D}_g$ and standard deviation of $0.2\bar{D}_g$ (Fig. \ref{RVE_mesh}).

 The microstructure of the RVE was periodic along the three directions of the space and periodic boundary conditions were applied to all the nodes on the boundary of the domain. The displacements of each pair of nodes A and B located on opposite surfaces of the domain were linked according to

 \begin{equation}
 \mathbf{u}_B-\mathbf{u}_A=(\mathbf{\bar{F}}-\mathbf{I})L
 \end{equation}	
 
 \noindent where $L$ is the length of the cubic domain. The far-field deformation gradient $\mathbf{\bar{F}}$ applied to the RVE is obtained by prescribing the displacements of three master nodes $M_i$.
 
 \begin{equation}
 \mathbf{u}(M_i)=(\mathbf{\bar{F}}-\mathbf{I})\mathbf{e}_i
 \end{equation}
 
 \noindent where $\mathbf{e}_i$ with $i=1,2,3$ stand for the orthogonal basis along the three Cartesian axes $x, y, z$. The master nodes are linked to the microstructure by three springs of negligible stiffness, joined to a fixed node inside the RVE. The displacement of the master nodes is prescribed by applying a nodal force $P_j$ to the master node $M_i$ and degree of freedom j according to
 
 \begin{equation}
 P_j(M_i)=(\bar{\bm{\sigma}}\mathbf{e}_i)_jA_i
 \end{equation}
 
 \noindent where $\bar{\bm{\sigma}}$ is the far-field stress tensor and $A_i$ is the projection of the current area of the face perpendicular to the $\mathbf{e_i}$ in this direction.

\begin{figure}[!t]
\centering
\includegraphics[width=0.65\textwidth]{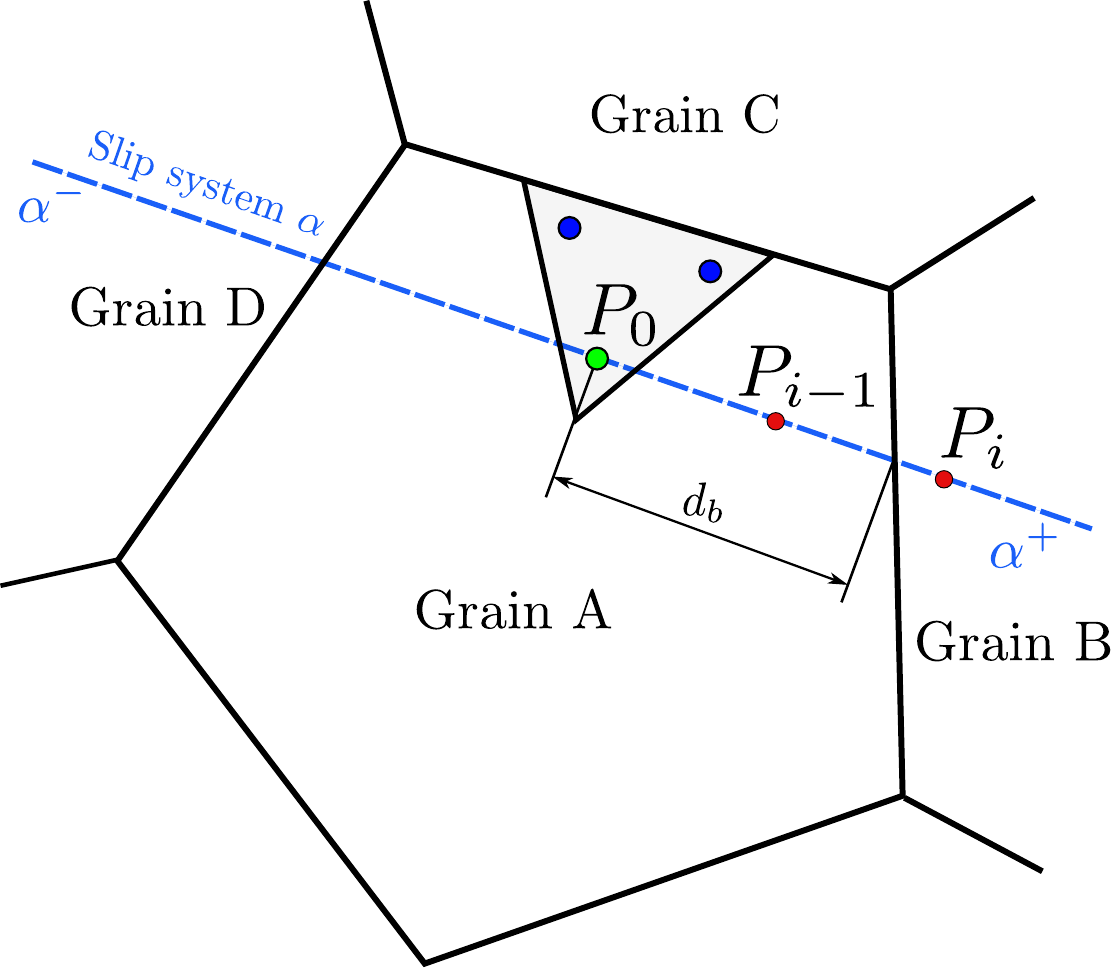}
\caption{Schematic of the calculation of the distance from a Gauss point to the nearest GB along one slip direction.}
\label{distance_calc_1}
\end{figure}

In order to apply eqs. (6) and (7), it is necessary to calculate the distance $d_b$ from each Gauss integration point to the nearest GB along the slip direction corresponding to each slip system and to identify the grain across the boundary. This latter information is necessary to apply the geometrical criteria ($LRB$, $m'_{\alpha\beta}$, $ \Delta b$ or any combination thereof) to assess whether slip transfer across the GB is possible for each slip system. This information only depends on the geometry of the RVE and the discretization and the values of $d_b$ for each slip system in each Gauss point are calculated and stored before the numerical analysis. The strategy to determine $d_b$ is schematically presented in Figure \ref{distance_calc_1}, which shows a simplified 2D representation of a section of a polycrystal in which grain A is surrounded by grains B, C and D. The integration point $P_0$ that belongs to a tetrahedral element of grain A (shaded in grey) is plotted in green in Fig. \ref{distance_calc_1} and the slip direction corresponding to the slip system $\alpha$ is plotted as a dashed blue line. The first step to calculate the distance $d_b$ is to identify positions $P_0$ to $P_1$ along the positive slip direction (indicated by $\alpha^+$ in the figure) using the distance 0.5$R_{eq}$, where $R_{eq}$ is the equivalent grain A radius (determined from the volume of the grain assuming that it is spherical). If $P_1$ belongs to grain A (which is easily determined from the coordinates of the vertices of grain A, which is a convex polyhedron), the process is repeated until the position of $P_i$ is outside of grain A and within grain B. Then, a bisection method is used to determine the position of the GB between $P_i$ and $P_{i-1}$ with a given tolerance and to determine $d_b$ along the positive slip direction. The same procedure is repeated for the negative slip direction and the corresponding distance to the opposite GB as well as the grain across this boundary is determined (grain D in this case). The minimum value of $d_b$ from both directions is stored as $d_b$ together with the identifier for the nearest neighbor grain. It was assumed that $d_b$ is constant throughout the analysis, a reasonable assumption because of the small applied far-field strain in the simulations (5\%).

An exception appears when the integration point is lying next to the outer boundary of the RVE because a point $P_i$ outside the RVE does not lie in the convex hull of any grain. In this case, the grain across the boundary along any slip system has to be determined taking into account the periodicity of the RVE. The point $P_i$ outside the RVE is shifted in the three directions of the space by a distance $\pm L$. Because of the periodicity of the RVE, one of the new positions of $P_i$ has to fall within one grain, which is the nearest neighbor along such slip system. 

Once the neighbor grain across the boundary has been identified for each slip system $\alpha$  at each Gauss integration point in grain A, the likelihood of slip transfer from the slip system $\alpha$ in grain A to any of the slip systems $\beta$ in the neighbor grain B can be assessed using any of the geometrical criteria detailed in the introduction. Two of them, based on the Luster-Morris geometric compatibility criterion $\mprime$ and on the residual Burgers vector $ \Delta b_{\alpha\beta}$, have been used in this investigation. They can be computed as \citep{Bayerschen2016a} (Fig. \ref{st_geometry}),

 \begin{equation}\label{GF}
m'_{\alpha\beta}=(\bm{n}_{\alpha}\cdot\bm{n}_{\beta})(\bm{b}_{\alpha}\cdot\bm{b}_{\beta}) \quad {\rm and} \quad \Delta b_{\alpha\beta} = |\bm{b}_{\alpha}-\bm{b}_{\beta}|
 \end{equation}

\noindent where $\bm{n}_{\alpha}$ and $\bm{n}_{\alpha}$ are unit vectors perpendicular to the slip plane $\alpha$ in grain A and slip plane $\beta$ in grain B. Obviously, all the vectors in eq. \eqref{GF} have to be expressed in the same reference frame. The orientation of grain A with respect to the Cartesian reference frame of the RVE is given by the Euler angles ($\varphi_1^A$, $\phi^A$, $\varphi_2^A$) while that of grain B is given by  ($\varphi_1^B$, $\phi^B$, $\varphi_2^B$). The vectors $\bm{n}'_{\alpha}$ and $\bm{b}'_{\alpha}$ expressed in the reference frame of crystal A can be transformed to the reference frame of crystal B, $\bm{n}_{\alpha}$ and $\bm{b}_{\alpha}$,  according to

 \begin{equation}\label{RM}
\bm{n}_{\alpha} = \bm{n}'_{\alpha} \bm{G}_A^{-1} \bm{G}_B  \quad {\rm and} \quad \bm{b}_{\alpha}=\bm{b}'_{\alpha}\bm{G}_A^{-1} \bm{G}_B \end{equation}

\noindent where $\bm{G}_A$ and $\bm{G}_B$ stnda for the orientation matrices of grains A and B, where, again, orientations are expressed with respect to the Cartesian reference frame of the RVE as a function of the Euler angles of each grain according to \citep{S2009}
\begin{equation*}
    \boldsymbol{G} =
    \begin{pmatrix}
        \cos\varphi_1 \cos\varphi_2 - \sin\varphi_1\sin\varphi_2\cos\phi &
        \sin\varphi_1\cos\varphi_2+\cos\varphi_1\sin\varphi_2\cos\phi
        & \sin\varphi_2\sin\phi \\
        -\cos\varphi_1\sin\varphi_2-\sin\varphi_1\cos\varphi_2\cos\phi & -\sin\varphi_1\sin\varphi_2+\cos\varphi_1\cos\varphi_2\cos\phi &\cos\varphi_2\sin\phi  \\
        \sin\varphi_1\sin\phi & -\cos\varphi_1\sin\phi& \cos\phi
    \end{pmatrix}.
\end{equation*}

Slip transfer between slip systems $\alpha$ and $\beta$ is allowed when $\mprime$ is above (or $\Delta b_{\alpha\beta}$ below) a threshold value. If the condition of slip transfer is fulfilled for any $\beta$ slip system in the neighbor grain, eq. (6) is used as the constitutive equation of slip system $\alpha$ at the Gauss integration point of the  crystal. Otherwise, slip transfer from $\alpha$ to $\beta$ is not possible due to the geometric incompatibility and the constitutive equation is expressed by eq. (7). 

It should be noted that these changes in the slip behavior due to the presence of impenetrable GBs are only active near to the grain boundaries (as dictated by the condition $K_s/d_b > 1/\sqrt{\ell^{\alpha}}$). Moreover, the geometrical analysis of slip transfer at each GB indicates how many pairs of slip systems  are suitably oriented to transfer slip across the boundary, leading to a classification of GBs from fully opaque (slip transfer is not possible for any pair of slip systems) to translucent (slip transfer is possible for some pairs of slip systems) to fully transparent (slip transfer is allowed for all 12 pairs of slip systems). 

The mechanical behavior of the RVEs under uniaxial tension was simulated using Abaqus/Standard \cite{Abaqus} within the framework of the finite deformations theory with the initial unstressed state as reference. The constitutive behavior of each crystal was governed by the crystal plasticity model presented in Section \ref{sec_cp_model}, that was implemented in Abaqus through a user-defined material model (UMAT). 


\section{Results and discussion}
\label{sec_results}

\subsection{Selection of RVE size and discretization}
\label{sec_rve_discretization}
An initial set of simulations was carried out in RVEs of Cu with fully opaque GBs to assess the effect of the discretization and of the number of grains in the RVE on the calculated stress-strain curves. RVEs with 100 grains were discretized with either 12000, 150000 and 350000 C3D10M elements. The coarse discretization overestimated the flow strength of the polycrystal by $\approx$ 10\% while the difference in flow strength between the medium and fine discretizations was only 2\%. In addition, the mechanical response of RVEs with 50, 100 and 200 grains and random texture was calculated using approximately the same discretization ($\approx$ 50000 elements). The differences in the flow strength between the RVEs with 50 and 100 grains were about 5\% and only of 2\% between the RVEs with 100 and 200 grains. Thus, RVEs with 100 grains discretized with 150000 C3D10M elements were selected for the analysis in order to optimize the balance between time and accuracy (Figure \ref{RVE_mesh}). Simulations of the same RVE with three different sets of random orientations showed differences of 4\% between the maximum and minimum flow stress. So, the texture that led to a flow stress in between the maximum and the minimum was selected for the simulations presented in this paper. This ensures that the flow stresses predicted by the different models were very close to the mean obtained by averaging the results of different RVEs and that the maximum differences from this mean value should be $<$ 2\% regardless of the particular random grain orientation distribution in the RVE.

\subsection{Effect of slip transfer on the flow strength}
\label{sec_resu_st}

The effect of grain size and grain boundary character on the mechanical response of Al and Cu polycrystals was analyzed using the crystal plasticity model and the computational homogenization scheme presented above. To this end, RVEs with four different grain sizes $\bar{D}_g$ (10, 20, 40 and 80 $\mu m$) and the same set of randomly-generated orientations were generated. Simulations were performed at quasi-static strain rates ($\approx10^{-3}$ $s^{-1}$) under uniaxial tension up to 5\% applied strain. The initial dislocation density was $\approx$ 10$^{12}$ $m^{-2}$, evenly distributed among the 12 slip systems, which corresponds to well-annealed polycrystals. The parameters of the crystal plasticity model for both Al and Cu can be found in Tables \ref{tab_FCC_parameters} and \ref{tab_single_crys}.

\begin{figure}[!t]
	\centering
	\includegraphics[width=0.94\textwidth]{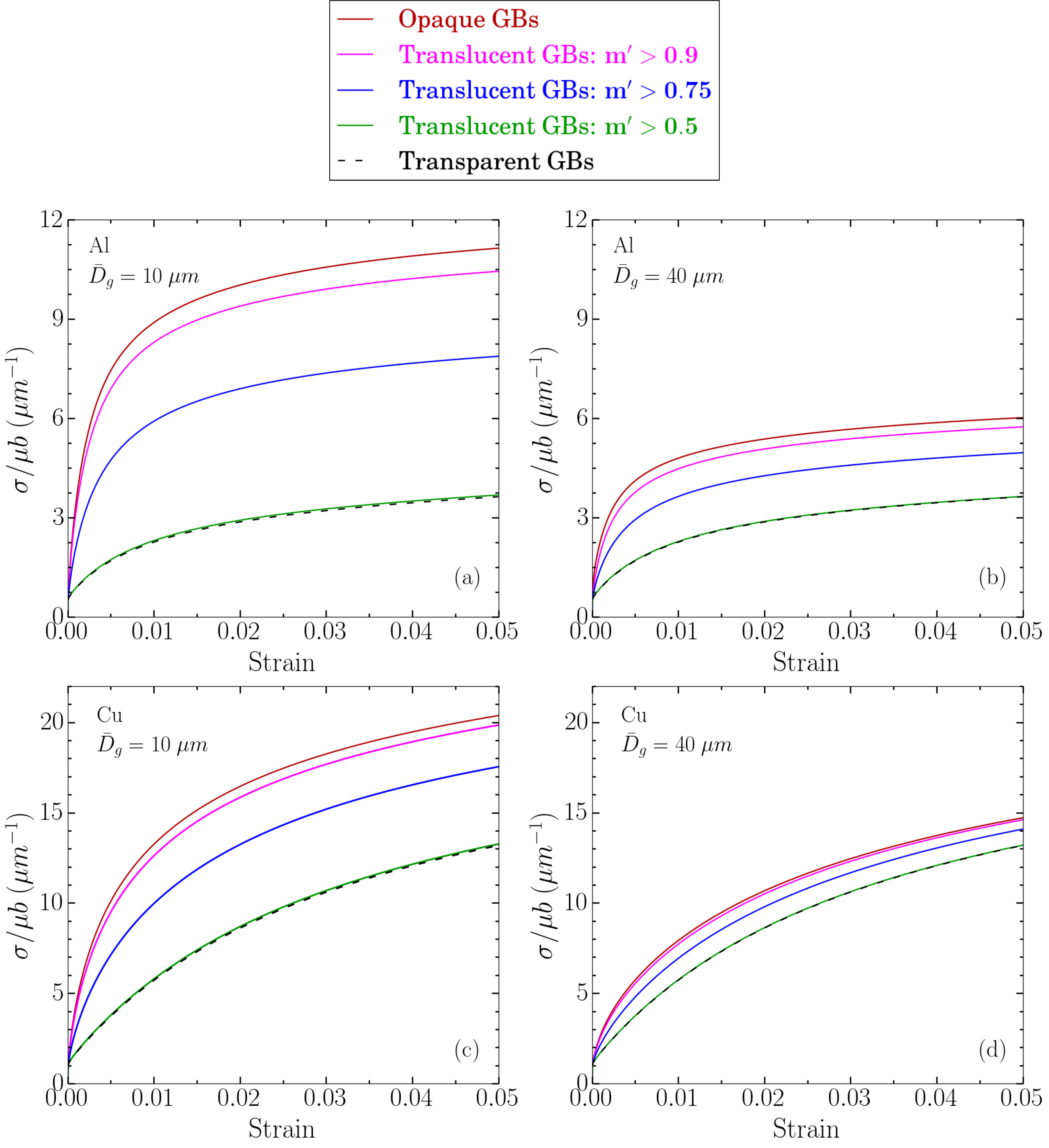}	
	\caption{Engineering stress-strain curves of Al and Cu polycrystals as a function of grain boundary type: (a) Al $\bar{D}_g$ =10 $\mu$m, (b) Al $\bar{D}_g$ = 40 $\mu$m, (c) Cu $\bar{D}_g$ = 10 $\mu$m and (d) Cu $\bar{D}_g$ = 40 $\mu$m.}
	\label{Fig_ss_AlCu_1040}
\end{figure}

\indent The effect of grain boundary character on the engineering stress-strain curves in tension of Al and Cu polycrystals with different average grain sizes ($\bar{D}_g$ = 10 $\mu$m and 40 $\mu$m) is presented in Figs. \ref{Fig_ss_AlCu_1040}a to d. The stress in these plots has been normalized by $\mu b$ to reveal the differences in strain hardening between Al and Cu as a result of the interaction, accumulation and annihilation of dislocations and to compensate for the effect of the differences in elastic constants. Five different curves are plotted in each figure: the strongest polycrystals correspond to simulations in which all GBs are assumed to be opaque and, thus, the constitutive equation of the material is given by eq. (7). In contrast, the softest response is given by the dashed black curves that were obtained assuming that all GBs were fully transparent. The constitutive equation of the polycrystal in this case is given by eq. (6) and the results of the simulations are independent of  $\bar{D}_g$ and stand for the behavior of a polycrystal with an "infinite" grain size. Obviously, the differences in the flow stress between  polycrystals with fully opaque and fully transparent GBs increase as the average grain size decreases, in agreement with the Hall-Petch effect. It should be noted that the stress-strain curves of Al polycrystals with either opaque or transparent GBs in Figs. \ref{Fig_ss_AlCu_1040}a and b show very little strain hardening for strains $>$ 2\%  and this behavior is associated with the high values of $K$ and $y_c$ in the constitutive model. Dislocation multiplication is responsible for the strain hardening in the bulk crystals, and decreases as the similitude coefficient $K$ increases because the dislocation MFP is proportional to $K$. Moreover, dislocation annihilation -which reduces the strain hardening- is more efficient for higher $y_c$. in contrast, the stress-strain curves of Cu polycrystals present continuous strain hardening because $K$ and $y_c$ are smaller. Thus,  Al and Cu represent two FCC polycrystals with very different strain hardening properties.

\begin{figure}[t!]
	\centering
	\includegraphics[width=\textwidth]{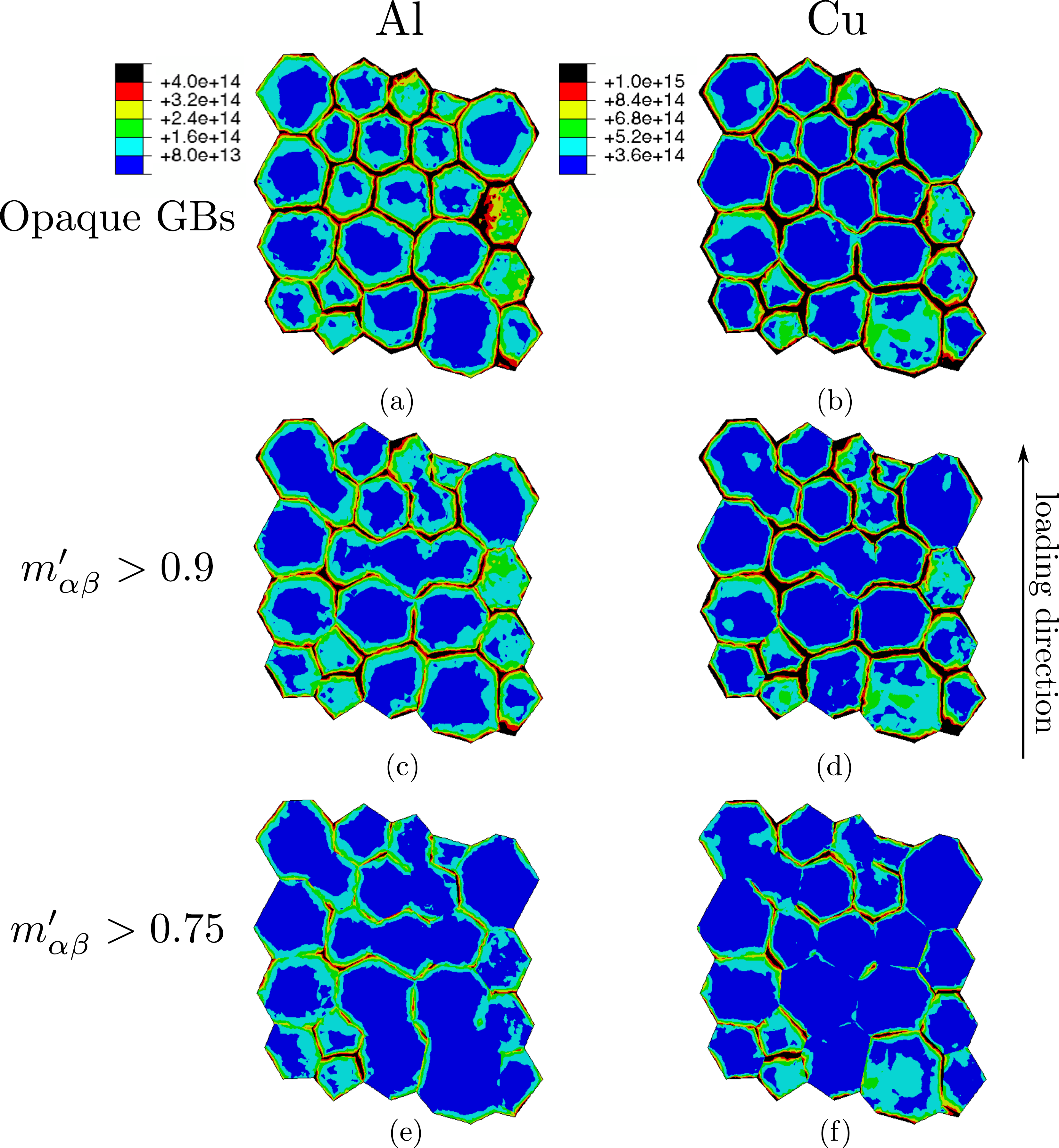}
	\caption{Spatial distribution of the total dislocation density (in $m^{-2}$) in a cross-section of the RVE of the Al and Cu polycrystals with an average  grain size of 10 $\mu$m deformed up to 5\%. (a) Al, opaque GBs. (b) Cu, opaque GBs. (c) Al, translucent GBs with $\mprime>0.9$. (d) Cu, translucent GBs with $\mprime>0.9$. (e) Al, translucent GBs with $\mprime>0.75$. (f) Cu, translucent GBs with $\mprime>0.75$.}
	\label{Fig_contour_AlCu_dd}
\end{figure}

The three curves between those corresponding to fully opaque and fully transparent  GBs in Figs. \ref{Fig_ss_AlCu_1040}a to d show the behavior of polycrystals where slip transfer is allowed between slip systems that intersect at a GB when the Luster-Morris parameter $\mprime$ is higher than a threshold given by 0.9, 0.75 or 0.5. The higher  the threshold, the closer the stress-strain curves are to the fully-opaque case. On the other hand, the stress-strain curves obtained with a threshold $\mprime > 0.5$  are almost the same as those obtained with fully-transparent GBs for both Al and Cu. The effect of slip transfer on the formation of dislocation pile-ups near the GBs can be ascertained from the spatial distribution plots of the dislocation density in a cross-section of the RVEs corresponding to Al (Figs. \ref{Fig_contour_AlCu_dd}a, c and e) and Cu polycrystals (Fig. \ref{Fig_contour_AlCu_dd}b, d and f) with an average grain size of 10 $\mu$m deformed up to 5\%. Obviously, high dislocation densities are found  at all of GBs if they are assumed to be opaque (Figs. \ref{Fig_contour_AlCu_dd}a and b), leading to a large increase in the flow stress observed in Figs.  \ref{Fig_ss_AlCu_1040}a and c for Al and Cu polycrystals with $\bar{D}_g$ =10 $\mu$m. As slip transfer is allowed, some opaque GBs become translucent or transparent as slip transfer is allowed, as shown in Figs. \ref{Fig_contour_AlCu_dd}c and e for Al and in d and e for Cu, and the number of GBs in which dislocation pile-ups have disappeared increases as the geometrical threshold for slip transfer is reduced in the model. In the case of $\mprime > 0.5$ (not shown in the figure), practically all GBs are transparent and no dislocation pile-ups can be found.

\begin{figure}[t!]
	\centering
	\includegraphics[width=\textwidth]{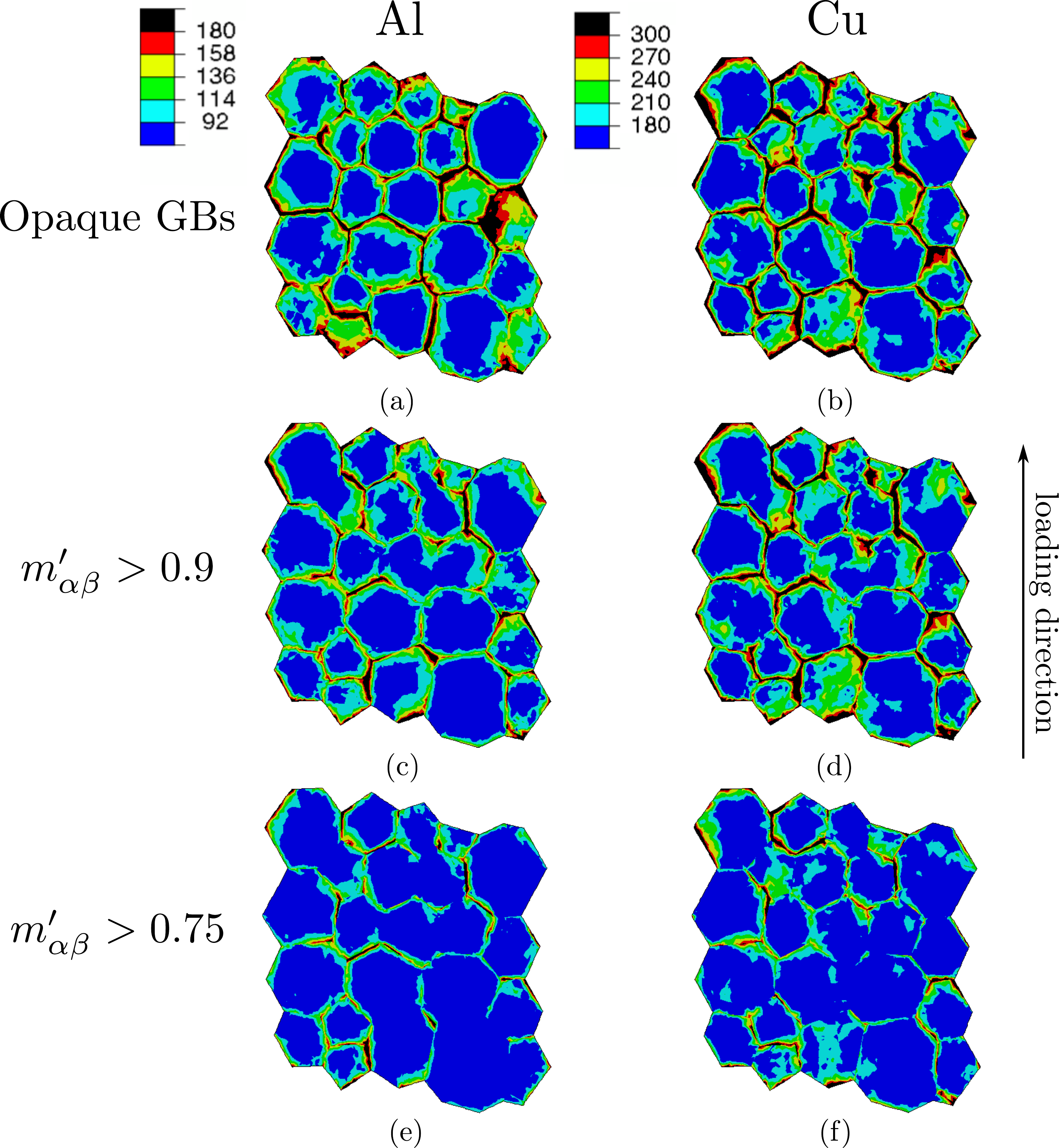}
	\caption{Spatial distribution of the Von Mises stress (in MPa) in a cross-section of the RVE of the Al and Cu polycrystals with an average grain size of 10 $\mu$m grain size deformed up to 5\%. (a) Al, opaque GBs. (b) Cu, opaque GBs. (c) Al, translucent GBs with $\mprime>0.9$. (d) Cu, translucent GBs with $\mprime>0.9$. (e) Al, translucent GBs with $\mprime>0.75$. (f) Cu, translucent GBs with $\mprime>0.75$.}
	\label{Fig_contour_AlCu_svm}
\end{figure}

The formation of pile-ups at GBs is directly associated with an increase in the local stresses, as shown in the spatial distribution plots of the Von Mises stress in the cross-section of the RVEs of Al (Fig. \ref{Fig_contour_AlCu_svm}a, c and e) and Cu polycrystals (Fig. \ref{Fig_contour_AlCu_svm}b, d and f) with an average grain size of 10 $\mu$m deformed up to 5\%. The more transparent the GB, the lower the stress concentration and, thus, slip transfer is likely to play a dominant role on the determination of the GBs at which damage is likely to develop in polycrystals.

The results presented above show how slip transfer influences the local dislocation density and flow stress at GBs. Obviously, the reduction in the strength of the polycrystal should be related to the threshold for translucent GBs in the RVE. This information can be obtained from the geometrical analysis to determine the nearest neighbor grain for each slip system $\alpha$ presented in section \ref{sec_ch_framework} and the corresponding $\mprime$ values for all the $\beta$ slip systems in the nearest neighbor grain. Then, a GB between grains A and B is characterized from the viewpoint of slip transfer by all the $\mprime$ values between all the slip systems in grains A and B.  The translucency of the grain boundary can be assessed by analyzing the number of slip systems that fulfil a  slip transfer criterion. Taking into account that FCC crystals have 12 slip systems from the $\{1 1 1\}<1 1 0>$ family, a $12\times12$ $\mprime$ matrix is calculated for pairs of neighbor grains, where each row relates the geometric compatibility of the lattice  from a given incoming slip system $\alpha$. In this context, slip transfer is considered possible along a certain slip system, for instance $\alpha=1$, when any of the  $m'_{1\beta}$ values for the 12 possible outgoing slip systems $\beta$ in the nearest neighbor grain is above the  threshold for slip transfer. Thus, the number of slip systems in grain A that can transfer slip  can be determined for each grain boundary and the grain boundaries in the RVE can be classified according to their translucency to dislocations in three distinct groups: opaque GBs (slip transfer is not allowed for any pair of slip systems), translucent or partially-transparent GBs (slip transfer is  allowed along several pairs of slip systems) and fully transparent GBs (slip transfer is allowed in all slip systems).

\begin{figure}[t!]
	\centering
	\includegraphics[width=\textwidth]{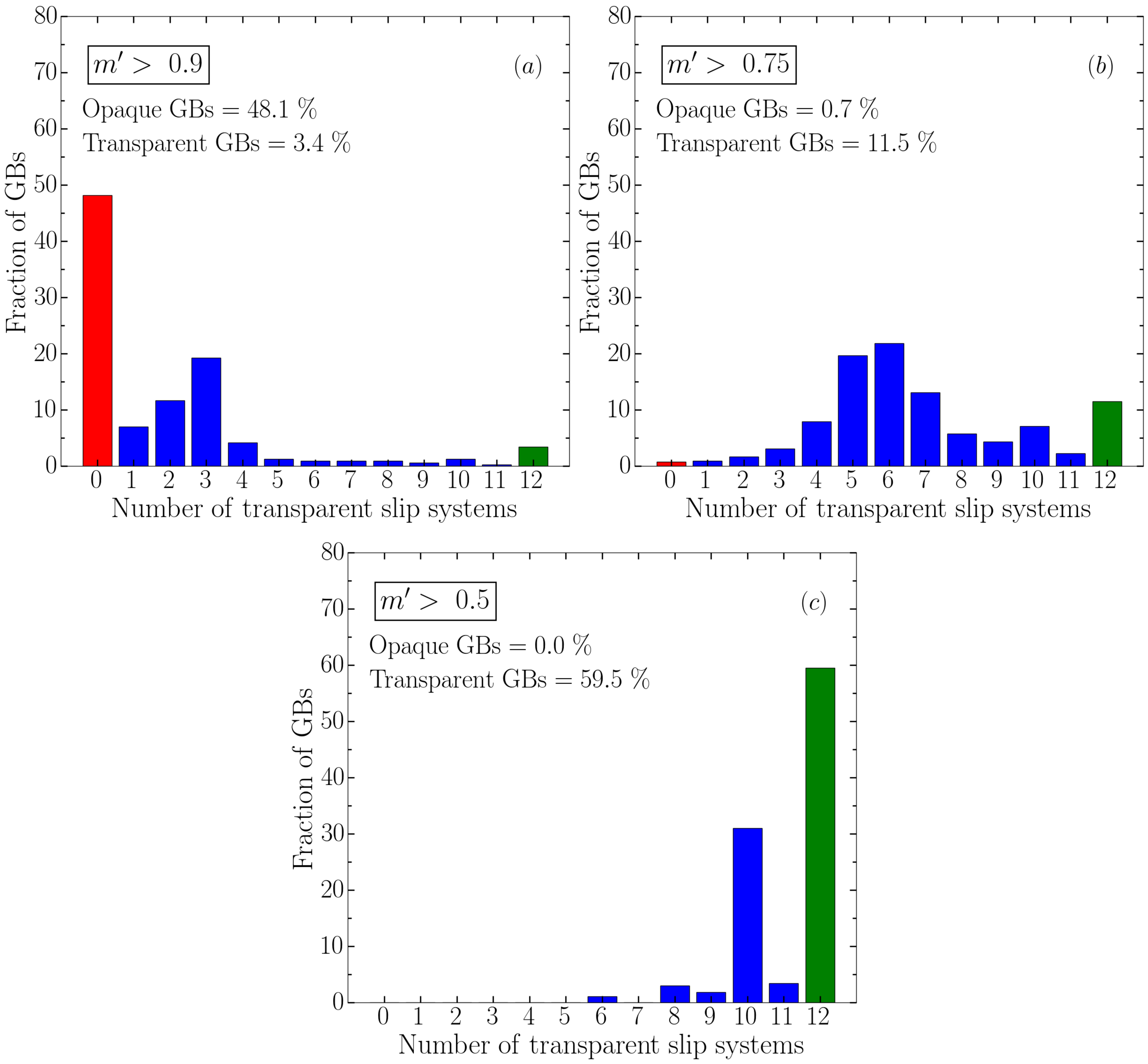}
	\caption{Fraction of GBs in the RVE as a function of the number of transparent slip systems. (a) $\mprime>0.9$. (b) $\mprime>0.75$. (c) $\mprime>0.5$.}
	\label{Fig_hist_st_mp}
\end{figure}

The fraction of GBs in the RVE as a function of the number of transparent slip systems across the boundary is plotted in Figs. \ref{Fig_hist_st_mp}a to c for different values of the threshold $\mprime$ for slip transfer. If  the threshold for slip transfer is very high, $\mprime >$ 0.9 (Fig. \ref{Fig_hist_st_mp}a), 48\% of the GBs are fully opaque while only 3.4\% of the GB are fully transparent (all slip systems in one grain can transfer slip to -at least- one slip system in the neighbor grain). Most of the remaining slip systems are translucent and contain 1 to 4 slip systems that can transfer slip to the neighbor grain. The number of GBs where 5 to 11 slip systems  can transfer slip is  negligible. When the threshold for slip transfer is reduced to $\mprime >$ 0.75, the fraction of fully opaque GBs drops to $<$ 1\% while the number of fully transparent GBs increases to 11.5\%. Moreover, most of the GBs can transfer slip across the GB along 4 or more slip systems. Finally, if the threshold for slip transfer is further reduced to $\mprime >$ 0.5 (Fig. \ref{Fig_hist_st_mp}c), almost 60\% of the GBs are fully transparent and the remaining ones can transfer slip along many slip systems, leading to a polycrystal in which the effect of GBs on the strength can be neglected.

\begin{figure}[!t]
	\centering
	\includegraphics[width=\textwidth]{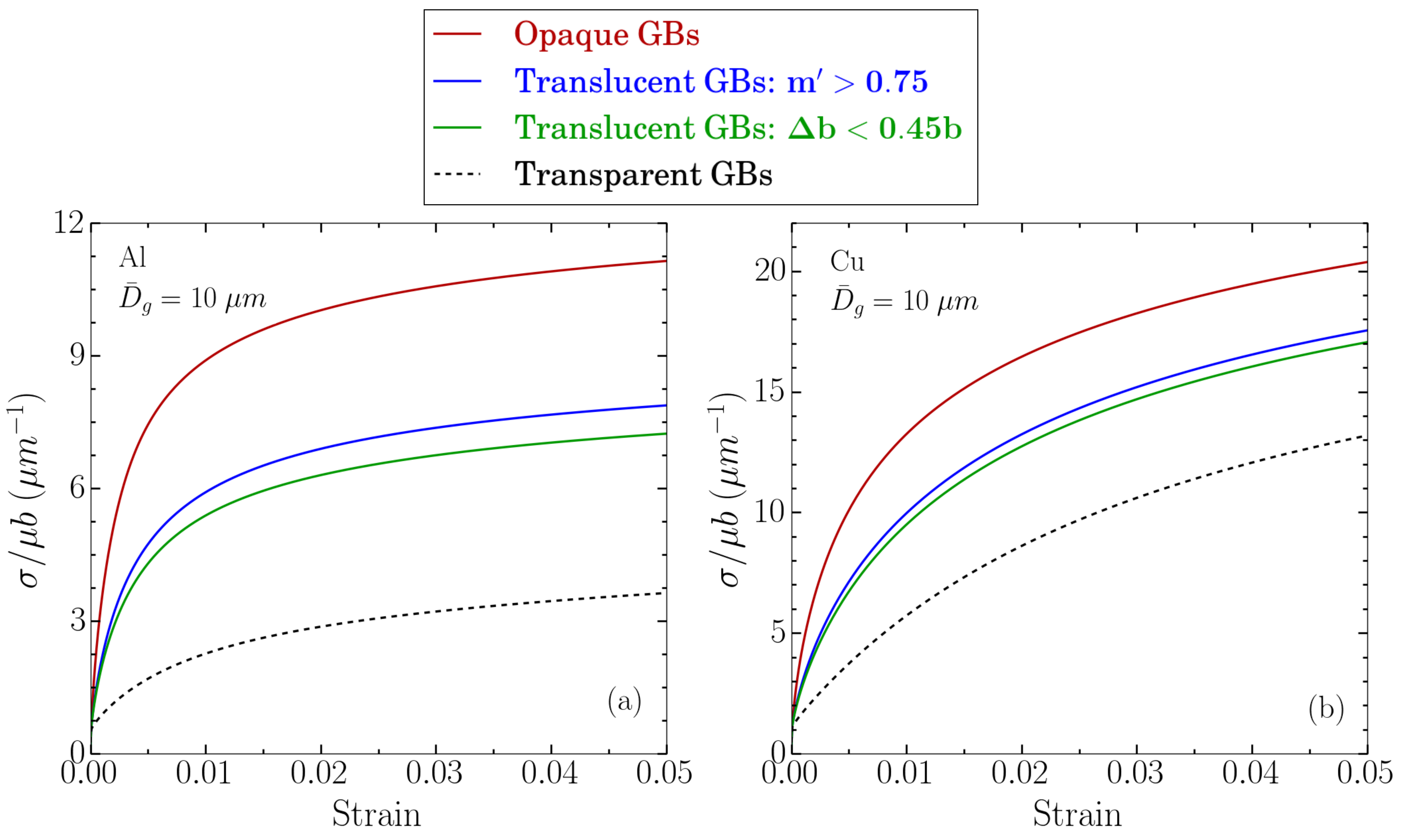}
	\caption{Engineering stress-strain curves of polycrystals with an average grain size of  10 $\mu$m for  opaque and transparent GBs as well as  slip transfer criteria of $\mprime>0.75$ or $\Delta b_{\alpha\beta} < 0.45b$. (a) Al. (b) Cu.}
	\label{Fig_ss_mp_rb}
\end{figure}

\begin{figure}[t!]
	\centering
	\includegraphics[width=0.98\textwidth]{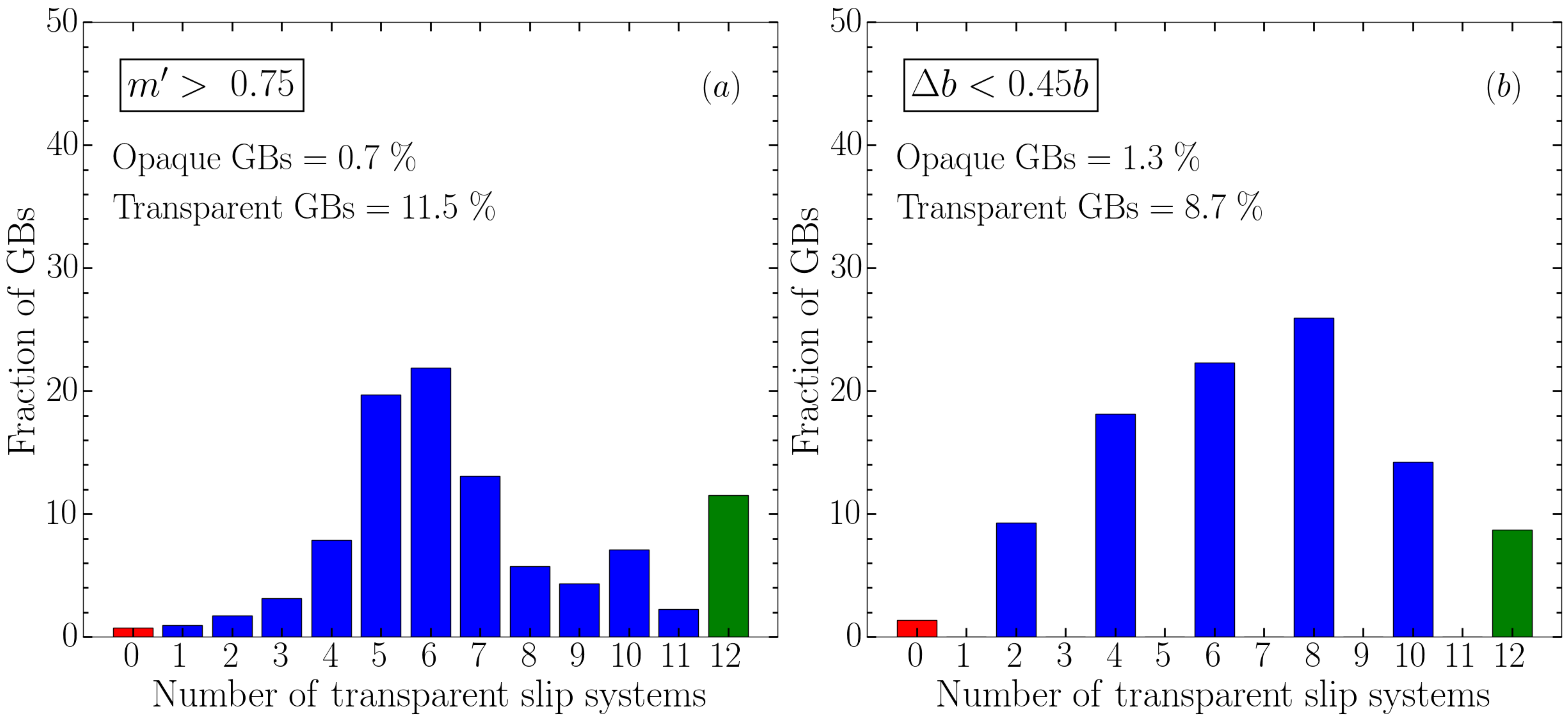}
	\caption{Fraction of  GBs in the RVE as a function of the number of transparent slip systems. (a) $\mprime>0.75$. (b) $\Delta b_{\alpha\beta} < 0.45b$.}
	\label{Fig_hist_mp_rb}
\end{figure}

So far, slip transfer in the simulations has been characterized by the Luster-Morris compatibility factor  but any other geometrical factor can be used in the model. For instance, the engineering stress-strain curves  for Al and Cu polycrystals with an average grain size of 10 $\mu$m are plotted in Figs. \ref{Fig_ss_mp_rb}a and b, respectively, when slip transfer across the GB was allowed when the residual Burgers vector $\Delta b_{\alpha\beta} < 0.45b$. The curves corresponding to fully opaque and fully transparent GBs (as well as those obtained with a slip transfer threshold defined by $\mprime >$ 0.75) are also plotted for comparison. It is worth noting that the stress-strain curves corresponding to $\mprime >$ 0.75 and $\Delta b_{\alpha\beta} < 0.45b$ are almost the same for both Al and Cu, suggesting that the reduction in the flow stress due to slip transfer is mainly controlled by the fraction of opaque, translucent and transparent GBs and not by the particular criterion used. This assumption is confirmed by the histograms of the different types of GBs in the RVE from the viewpoint of slip transfer plotted in Fig. \ref{Fig_hist_mp_rb} for both slip transfer thresholds. The fraction of fully opaque and fully transparent GBs is very similar in both cases and most of the GBs are translucent. Thus,  slip transfer is allowed in a wide number of slip systems (from 2 to 10) for both slip transfer thresholds. It is interesting to notice that pairs of slip systems of an FCC crystal share the same Burgers vectors and, thus, the number of transparent or opaque slip systems in the case of the residual Burgers vector criterion is always an even number. 

\subsection{Effect of slip transfer on the Hall-Petch law}
\label{sec_resu_st}

\begin{figure}[t!]
	\centering
	\includegraphics[width=\textwidth]{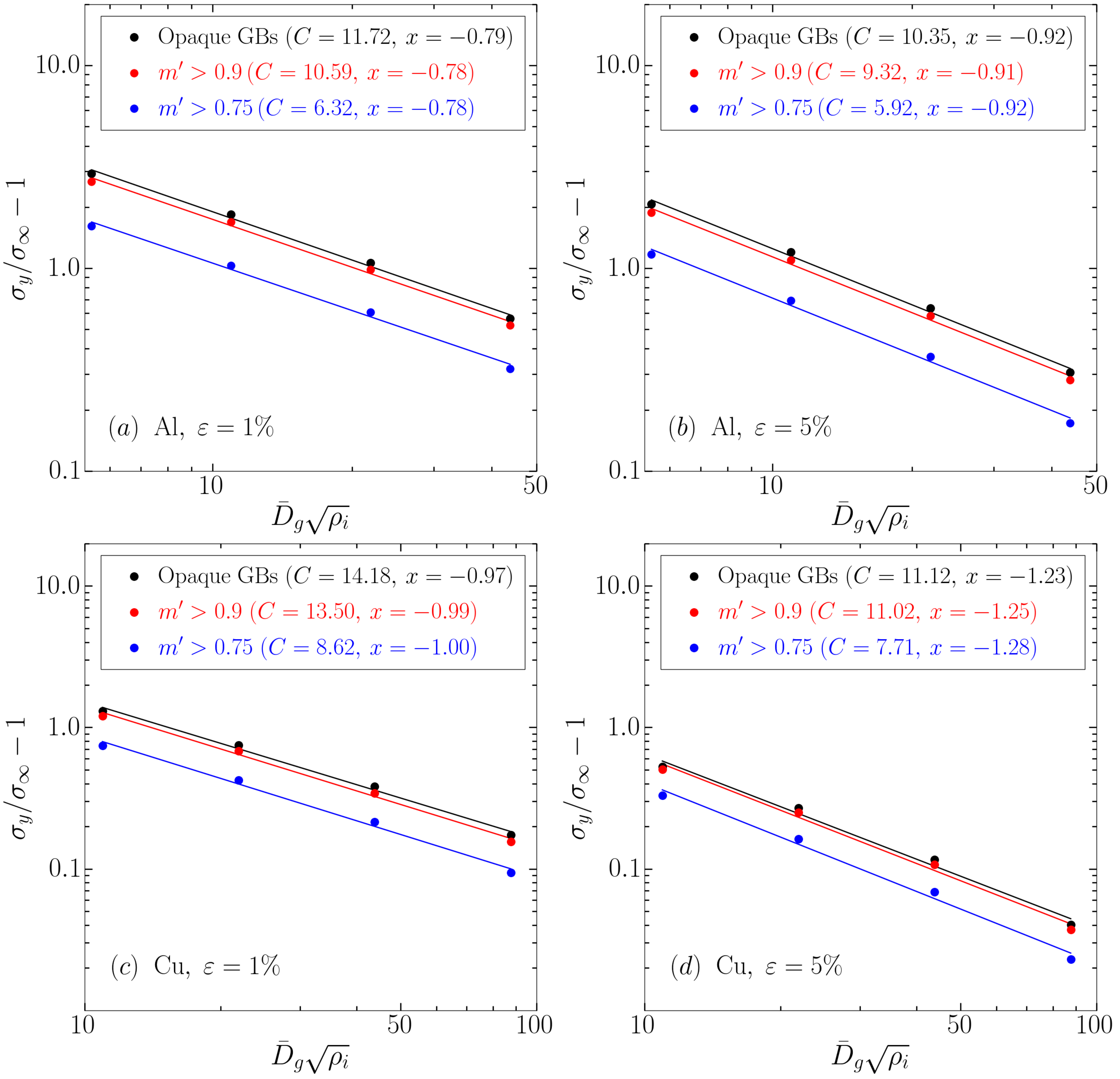}
	\caption{Grain boundary strengthening in Al and Cu polycrystals as a function of the adimensional parameter $\bar{D}_g\sqrt{\rho_i}$. (a) Al, $\varepsilon=1\%$. (b) Al, $\varepsilon=5\%$. (c) Cu, $\varepsilon=1\%$. (d) Cu, $\varepsilon=5\%$. The results of the simulations with fully opaque GBs and with thresholds $\mprime>0.9$ and $\mprime>0.75$ for slip transfer are plotted in each figure.}
	\label{Fig_effect_GB}
\end{figure}

The experimental evidence indicates that the strength of polycrystals increases as the grain size decreases following a power-law of the grain size.  Based on theoretical results \citep{zaiser2014scaling} and on dislocation dynamics simulations \citep{el2015unravelling}, the strengthening caused by GBs, $\sigma_y - \sigma_{\infty}$, has been proposed to scale with the average grain size $\bar{D}_g$ and the square root of the initial dislocation density in the polycrystal $\sqrt{\rho_i}$ according to \citep{Haouala2018}

\begin{equation}
\sigma_y/\sigma_{\infty}-1 = C(\bar{D}_g\sqrt{\rho_i})^{-x}
\label{eq_str_GB}
\end{equation}

\noindent where $\sigma_y$ is the flow strength of the poycrystal, $\sigma_{\infty}$ the flow stress of a polycrystal with "infinite" grain size and $C$ and $x$ are material constants that depend on the physical parameters of the model for each FCC metal, mainly the similitude constant $K$ and the effective annihilation distance $y_c$. Eq. \eqref{eq_str_GB} was in very good agreement with results of the simulations for Al, Cu, Ni and Ag FCC polycrystals under the assumption that all GBs were opaque \citep{Rubio2019} leading to similar values of the exponent (in the range 0.7 to 0.9) when the applied strain was 1\%. The effect of slip transfer on the flow strength of Al and Cu polycrystals with different grain size (in the range 10 $\mu$m to 80 $\mu$m) deformed up to $\varepsilon$ =1\% and 5\% is plotted in Fig \ref{Fig_effect_GB}. The initial dislocation density in all cases was of the order of $10^{12}$ $m^{-2}$. 
The results of the simulations show that slip transfer reduces the strengthening provided by GBs but does not modify the exponent of eq. \eqref{eq_str_GB} when values of the threshold $\mprime$ for slip transfer are similar to those reported experimentally \citep{Hemery2018, alizadeh2020criterion, ZBL20}. The exponent depends on the FCC metal and on the applied strain. It is slightly lower (more negative) for Cu than for Al  and decreases in both cases when the strain to determine the flow strength of the polycrystal is increased from 1\% to 5\% due to the effect of annihilation of dislocations near the GBs where the dislocation density is maximum.  Moreover, the reduction in GB strengthening for Al is always higher than that for Cu when the thresholds for slip transfer are set to $\mprime>0.9$ or $\mprime>0.75$ although the differences are not huge. They can be attributed to changes in the dislocation accumulation and annihilation of dislocations near the GBs as a result of the differences in the physical parameters that control these phenomena in each metal.

\subsection{Comparison with experiments}
\label{sec_resu_experiments}

In order to assess the validity of the modelling strategy presented above, numerical simulations of the flow stress as a function of the grain size were carried out for Al, Cu, Ni, and Ag polycrystals and compared with experimental data in the literature \citep{Hansen1977, narutani1991grain, carreker1957tensile, hansen1982strain} . The experiments  in the literature were carried out at  room temperature under quasi-static loading conditions and, therefore, the effect of the strain rate on the flow stress was considered negligible. Due to the lack of information about the initial experimental dislocation density, it was assumed to be of the order of $10^{12}$ $m^{-2}$ in all cases, which is a typical value for well-annealed polycrystals. The parameters of the crystal plasticity model for Al, Cu, Ni, and Ag were obtained by \cite{Rubio2019} and can be found in Tables  \ref{tab_FCC_parameters} and \ref{tab_single_crys}. All the simulations were carried out using the RVE shown in Fig. \ref{RVE_mesh} with random texture and average grain sizes in the range 10 $\mu$m to 80 $\mu$m.

\begin{figure}[t!]
	\centering
	\includegraphics[width=\textwidth]{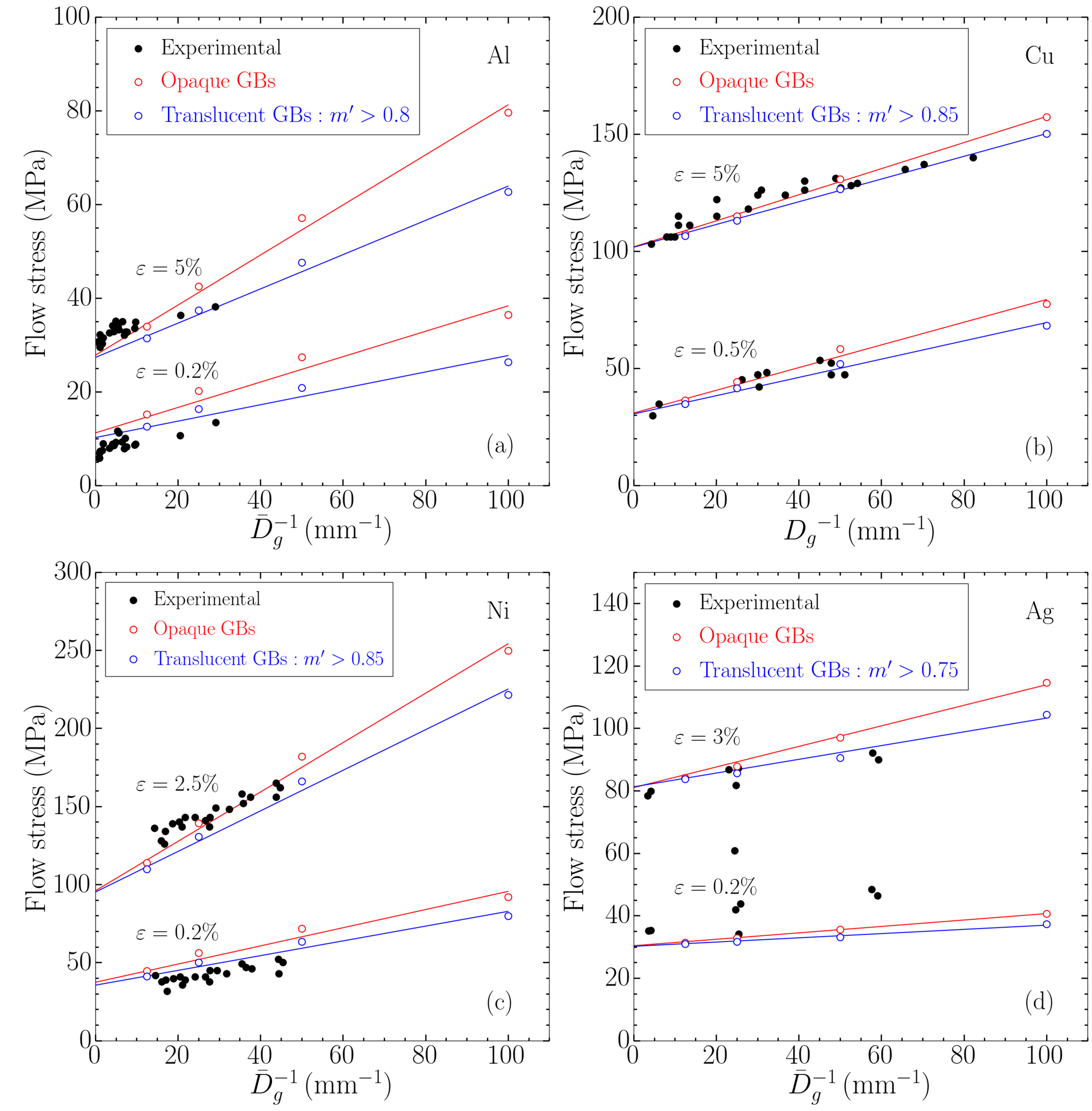}
	\caption{Experimental data from the literature and simulation results of the flow stress of Al, Cu, Ni and Ag polycrystals as a function of $\bar{D}_g^{-1}$ for different applied strains. (a) Al. (b) Cu. (c) Ni. (d) Ag. Experimental results are indicated by black circles, while simulation results without and with slip transfer are represented by open red and blue circles.}
	\label{Fig_flow_stress}
\end{figure}

The experimental results of the flow stress in tension are plotted as a function of the inverse of the average grain size, $\bar{D}_g^{-1}$ in Figs. \ref{Fig_flow_stress}a, b, c and d for Al, Cu, Ni, and Ag polycrystals, respectively. Data for two different strain values are shown in each figure. Two simulation results assuming fully opaque GBs (open red circles)  and translucent GBs  with a threshold in the Luster-Morris parameter (open blue circles) are compared. The thresholds selected for $\mprime$ were chosen for each alloy to get a best fit of the experimental data because accurate experimental information is not available. Nevertheless, they are in the range of the values reported by different experimental investigations of slip transfer \citep{Hemery2018, alizadeh2020criterion, ZBL20}. More importantly, good agreement between experiments and simulations is found when the average grain size is large ($\bar{D}_g^{-1} <$ 25 mm$^{-1}$, $\bar{D}_g >$ 40 $\mu$m), regardless of the inclusion of slip transfer in the simulations, but the trend of the experimental data follows the predictions of the simulations including slip transfer when the grain sizes are smaller than 20 $\mu$m  ($\bar{D}_g^{-1} >$ 50 mm$^{-1}$). It should be noted, however, that the experimental dataset of Ag is smaller than those of Al, Cu and Ni, and presents larger scatter in the flow stress values, especially for 0.2\% applied strain. For this reason, together with the absence of information about the initial dislocation density, it is not possible to conclude that the predictions with translucent GBs are more accurate than those carried out with opaque GBs for Ag at small strains.

It is also important to notice that slip transfer not only influences the overall strength of the polycrystal but also leads to dramatic changes in the stresses at the GBs as a function of the GB orientation (Fig. \ref{Fig_contour_AlCu_svm}). These differences in GBs as a function of orientation are likely to play a dominant role on the nucleation of damage by GB cracking  during monotonic and cyclic deformation or in aggressive environments \citep{MCR15, LHC21} as well as on the development of GB sliding during creep deformation \citep{WA04}. It is obvious that the combination of the crystal plasticity model presented in this paper together with cohesive elements at the GBs \citep{ABS18} would provide a novel way to explore the role of opaque, translucent and transparent GBs on the ductility, formability and forming limits of polycrystalline materials.

\section{Conclusions}
\label{sec_conclusion}

The effect slip transfer on the flow strength of  FCC polycrystals has been analyzed by means of full-field computational homogenization in combination with a physically-based crystal plasticity model.  The critical resolved shear stress for plastic slip in each slip system follows the modified Taylor model while the generation and annihilation of dislocations in each slip system in the bulk during deformation is described by a Kocks-Mecking law. This law is modified near the grain boundaries to discriminate between slip systems that can or cannot transfer slip across the boundary using geometrical criteria based on the orientation and activity of the slip systems of both grains near the boundary. This model leads to a classification of the grain boundaries as fully opaque (slip transfer is not possible for any slip system), fully transparent (slip transfer is possible for all slip systems) and translucent (slip transfer is possible for several slip systems).

The mechanical behavior of Al and Cu polycrystals with grain sizes in the range 10 $\mu$m to 80 $\mu$m and random texture was determined assuming that all grain boundaries were opaque, transparent or that slip transfer was possible for different thresholds of the  Luster-Morris geometric compatibility parameter $\mprime$ or of the residual Burgers vector $ \Delta b_{\alpha\beta}$. Slip transfer between neighbor grains suitably oriented led to a clear reduction in the flow stress of the polycrystals (as compared with the simulations with opaque grain boundaries) which was dependent on the fraction of translucent and transparent grain boundaries in the microstructure, as given  by the threshold in the slip transfer geometrical criterion. Moreover, dislocation densities and Von Mises stresses were much higher around opaque grain boundaries, which are potential places for damage nucleation.

The strengthening provided by grain boundaries in Al and Cu polycrystals was well described by $C(\bar{D}_g\sqrt{\rho_i})^{-x}$ where $\bar{D}_g$ is the average grain size and $\rho_i$ the initial dislocation density. $C$ depended on the material and on the fraction of translucent and transparent boundaries while $x$ was a function of the material and of the applied strain but was independent of slip transfer at grain boundaries.

Finally, the predictions of the simulations were compared with experimental data in the literature of the effect of grain size on the strength of Al, Cu, Ni and Ag polycrystals. It was found that the inclusion of slip transfer in the model led to more accurate predictions of the Hall-Petch effect particularly for small grain sizes ($<$ 20 $\mu$m).

\section*{Acknowledgements}
This work was funded by the European Research Council Advanced Grant VIRMETAL under the European Union's Horizon 2020 research and innovation programme (Grant agreement No. 669141) and by the HexaGB project of the Spanish Ministry of Science (reference RTI2018-098245).



\begin{thebibliography}{57}
\expandafter\ifx\csname natexlab\endcsname\relax\def\natexlab#1{#1}\fi
\expandafter\ifx\csname url\endcsname\relax
  \def\url#1{\texttt{#1}}\fi
\expandafter\ifx\csname urlprefix\endcsname\relax\def\urlprefix{URL }\fi

\bibitem[{Abaqus(2020)}]{Abaqus}
Abaqus, 2020. Analysis {U}ser's manual. Dassault Syst\`emes.

\bibitem[{Abuzaid et~al.(2016)Abuzaid, Sehitoglu, and Lambros}]{Abuzaid2016}
Abuzaid, W.~Z., Sehitoglu, H., Lambros, J., 2016. {Localisation of plastic
  strain at the microstructurlal level in Hastelloy X subjected to monotonic,
  fatigue, and creep loading: the role of grain boundaries and slip
  transmission}. Materials at High Temperatures 33~(4-5), 384--400.

\bibitem[{Acharya and Beaudoin(2000)}]{AB00}
Acharya, A., Beaudoin, A.~J., 2000. Grain size effects in viscoplastic
  polycrystals at moderate strains. Journal of the Mechanics and Physics of
  Solids 48, 2213--2230.

\bibitem[{Alabort et~al.(2018)Alabort, Barba, Sulzer, Libner, Petrinic, and
  Reed}]{ABS18}
Alabort, E., Barba, D., Sulzer, S., Libner, M., Petrinic, N., Reed, R., 2018.
  Grain boundary properties of a nickel-based superalloy: Characterisation and
  modelling. Acta Materialia 151, 377 -- 394.

\bibitem[{Alizadeh et~al.(2020)Alizadeh, Pe{\~{n}}a-Ortega, Bieler, and
  LLorca}]{alizadeh2020criterion}
Alizadeh, R., Pe{\~{n}}a-Ortega, M., Bieler, T.~R., LLorca, J., 2020. {A
  criterion for slip transfer at grain boundaries in Al}. Scripta Materialia
  178, 408--412.

\bibitem[{Ashby(1970)}]{Ashby1970}
Ashby, M.~F., 1970. {The deformation of plastically non-homogeneous materials}.
  Philosophical Magazine 21~(170), 399--424.

\bibitem[{Bargmann et~al.(2010)Bargmann, Ekh, Runesson, and Svendsen}]{BER10}
Bargmann, S., Ekh, M., Runesson, K., Svendsen, B., 2010. Modeling of
  polycrystals with gradient crystal plasticity: A comparison of strategies.
  Philosophical Magazine 90, 1263--1288.

\bibitem[{Bayerschen et~al.(2016)Bayerschen, McBride, Reddy, and
  B{\"{o}}hlke}]{Bayerschen2016a}
Bayerschen, E., McBride, A.~T., Reddy, B.~D., B{\"{o}}hlke, T., 2016. {Review
  on slip transmission criteria in experiments and crystal plasticity models}.
  Journal of Materials Science 51, 2243--2258.

\bibitem[{Bayley et~al.(2007)Bayley, Brekelmans, and Geers}]{BBG07}
Bayley, C.~J., Brekelmans, W. A.~M., Geers, M. G.~D., 2007. A three-dimensional
  dislocation field crystal plasticity approach applied to miniaturized
  structures. Philosophical Magazine 87, 1361--1378.

\bibitem[{Bertin et~al.(2013)Bertin, Capolungo, and
  Beyerlein}]{bertin2013hybrid}
Bertin, N., Capolungo, L., Beyerlein, I.~J., 2013. {Hybrid dislocation dynamics
  based strain hardening constitutive model}. International Journal of
  Plasticity 49, 119--144.

\bibitem[{Bieler et~al.(2019)Bieler, Alizadeh, Pe{\~{n}}a-Ortega, and
  Llorca}]{Bieler2019}
Bieler, T.~R., Alizadeh, R., Pe{\~{n}}a-Ortega, M., Llorca, J., 2019. {An
  analysis of (the lack of) slip transfer between near-cube oriented grains in
  pure Al}. International Journal of Plasticity 118, 269--290.

\bibitem[{Bieler et~al.(2014)Bieler, Eisenlohr, Zhang, Phukan, and
  Crimp}]{Bieler2014}
Bieler, T.~R., Eisenlohr, P., Zhang, C., Phukan, H.~J., Crimp, M.~A., 2014.
  {Grain boundaries and interfaces in slip transfer}. Current Opinion in Solid
  State and Materials Science 18~(4), 212--226.

\bibitem[{Carreker(1957)}]{carreker1957tensile}
Carreker, R.~P., 1957. {Tensile deformation of silver as a function of
  temperature, strain rate, and grain size}. JOM 9~(1), 112--115.

\bibitem[{Cheong et~al.(2005)Cheong, Busso, and Arsenlis}]{CBA05}
Cheong, K.~S., Busso, E.~P., Arsenlis, A., 2005. A study of microstructural
  length scale effects on the behavior of {FCC} polycrystals using strain
  gradient concepts. International {J}ournal of {P}lasticity 21, 1797--1814.

\bibitem[{{De Sansal} et~al.(2010){De Sansal}, Devincre, and
  Kubin}]{de2010grain}
{De Sansal}, C., Devincre, B., Kubin, L., 2010. {Grain size strengthening in
  microcrystalline copper: A three-dimensional dislocation dynamics
  simulation}. In: Key Engineering Materials. Vol. 423. Trans Tech Publ, pp.
  25--32.

\bibitem[{Devincre et~al.(2008)Devincre, Hoc, and
  Kubin}]{devincre2008dislocation}
Devincre, B., Hoc, T., Kubin, L., 2008. {Dislocation mean free paths and strain
  hardening of crystals}. Science 320~(5884), 1745--1748.

\bibitem[{Dunstan and Bushby(2013)}]{dunstan2013scaling}
Dunstan, D.~J., Bushby, A.~J., 2013. {The scaling exponent in the size effect
  of small scale plastic deformation}. International Journal of Plasticity 40,
  152--162.

\bibitem[{Dunstan and Bushby(2014)}]{dunstan2014grain}
Dunstan, D.~J., Bushby, A.~J., 2014. {Grain size dependence of the strength of
  metals: The Hall--Petch effect does not scale as the inverse square root of
  grain size}. International Journal of Plasticity 53, 56--65.

\bibitem[{El-Awady(2015)}]{el2015unravelling}
El-Awady, J.~A., 2015. {Unravelling the physics of size-dependent
  dislocation-mediated plasticity}. Nature Communications 6~(1), 1--9.

\bibitem[{Essmann and Mughrabi(1979)}]{essmann1979annihilation}
Essmann, U., Mughrabi, H., 1979. {Annihilation of dislocations during tensile
  and cyclic deformation and limits of dislocation densities}. Philosophical
  Magazine A 40, 731--756.

\bibitem[{Franciosi et~al.(1980)Franciosi, Berveiller, and
  Zaoui}]{franciosi1980latent}
Franciosi, P., Berveiller, M., Zaoui, A., 1980. {Latent hardening in copper and
  aluminium single crystals}. Acta Metallurgica 28~(3), 273--283.

\bibitem[{Friedman and Chrzan(1998)}]{friedman1998continuum}
Friedman, L.~H., Chrzan, D.~C., 1998. {Continuum analysis of dislocation
  pile-ups: Influence of sources}. Philosophical Magazine A 77, 1185--1204.

\bibitem[{Geuzaine and Remacle(2009)}]{Geuzaine2009}
Geuzaine, C., Remacle, J.~F., 2009. {Gmsh: A 3-D finite element mesh generator
  with built-in pre- and post-processing facilities}. International Journal for
  Numerical Methods in Engineering 79~(11), 1309--1331.

\bibitem[{Hall(1951)}]{hall1951deformation}
Hall, E.~O., 1951. {The deformation and ageing of mild steel: II
  Characteristics of the L{\"{u}}ders deformation}. Proceedings of the Physical
  Society. Section B 64~(9), 742--747.

\bibitem[{Hansen(1977)}]{Hansen1977}
Hansen, N., 1977. {The effect of grain size and strain on the tensile flow
  stress of aluminium at room temperature}. Acta Metallurgica 25~(8), 863--869.

\bibitem[{Hansen and Ralph(1982)}]{hansen1982strain}
Hansen, N., Ralph, B., 1982. {The strain and grain size dependence of the flow
  stress of copper}. Acta Metallurgica 30~(2), 411--417.

\bibitem[{Haouala et~al.(2020{\natexlab{a}})Haouala, Alizadeh, Bieler,
  Segurado, and LLorca}]{Haouala2019}
Haouala, S., Alizadeh, R., Bieler, T.~R., Segurado, J., LLorca, J.,
  2020{\natexlab{a}}. {Effect of slip transmission at grain boundaries in Al
  bicrystals}. International Journal of Plasticity 126, 102600.

\bibitem[{Haouala et~al.(2020{\natexlab{b}})Haouala, Lucarini, LLorca, and
  Segurado}]{Haouala2020}
Haouala, S., Lucarini, S., LLorca, J., Segurado, J., 2020{\natexlab{b}}.
  {Simulation of the Hall-Petch effect in FCC polycrystals by means of strain
  gradient crystal plasticity and FFT homogenization}. Journal of the Mechanics
  and Physics of Solids 134, 103755.

\bibitem[{Haouala et~al.(2018)Haouala, Segurado, and LLorca}]{Haouala2018}
Haouala, S., Segurado, J., LLorca, J., 2018. {An analysis of the influence of
  grain size on the strength of FCC polycrystals by means of computational
  homogenization}. Acta Materialia 148, 72--85.

\bibitem[{H{\'{e}}mery et~al.(2018)H{\'{e}}mery, Nizou, and
  Villechaise}]{Hemery2018}
H{\'{e}}mery, S., Nizou, P., Villechaise, P., 2018. {In situ SEM investigation
  of slip transfer in Ti-6Al-4V: Effect of applied stress}. Materials Science
  and Engineering A 709, 277--284.

\bibitem[{Hirth(1972)}]{Hirth1972}
Hirth, J.~P., 1972. {Influence of Grain Boundaries on Mechanical Properties.}
  Metall Trans 3~(12), 3047--3067.

\bibitem[{Hughes et~al.(2003)Hughes, Hansen, and
  Bammann}]{hughes2003geometrically}
Hughes, D.~A., Hansen, N., Bammann, D.~J., 2003. {Geometrically necessary
  boundaries, incidental dislocation boundaries and geometrically necessary
  dislocations}. Scripta Materialia 48~(2), 147--153.

\bibitem[{Kocks(1970)}]{kocks1970relation}
Kocks, U.~F., 1970. {The relation between polycrystal deformation and
  single-crystal deformation}. Metallurgical and Materials Transactions B
  1~(5), 1121--1143.

\bibitem[{Kocks et~al.(1979)Kocks, Argon, and Ashby}]{kocks1975thermodynamics}
Kocks, U.~F., Argon, A., Ashby, M.~F., 1979. {The Thermodynamics and Kinetics
  of Slip}. Progress in Materials Science 19, 1--291.

\bibitem[{Kocks and Mecking(2003)}]{kocks2003physics}
Kocks, U.~F., Mecking, H., 2003. {Physics and phenomenology of strain
  hardening: The FCC case}. Progress in Materials Science 48~(3), 171--273.

\bibitem[{Kubin(2013)}]{kubin2013dislocations}
Kubin, L., 2013. {Dislocations, Mesoscale Simulations and Plastic Flow}.
  Vol.~5. Oxford University Press.

\bibitem[{Lebensohn and Needleman(2016)}]{LN16}
Lebensohn, R.~A., Needleman, A., 2016. Numerical implementation of non-local
  polycrystal plasticity using {Fast Fourier Transforms}. Journal of the
  Mechanics and Physics of Solids 97, 333--351.

\bibitem[{Lee et~al.(1989)Lee, Robertson, and Birnbaum}]{Lee1989}
Lee, T.~C., Robertson, I.~M., Birnbaum, H.~K., 1989. {Prediction of slip
  transfer mechanisms across grain boundaries}. Scripta Metallurgica 23~(5),
  799--803.

\bibitem[{Li et~al.(2016)Li, Bushby, and Dunstan}]{li2016hall}
Li, Y., Bushby, A.~J., Dunstan, D.~J., 2016. {The Hall--Petch effect as a
  manifestation of the general size effect}. Proceedings of the Royal Society A
  472, 20150890.

\bibitem[{Liang et~al.(2021)Liang, Hure, Courcelle, Shawish, and
  Tanguy}]{LHC21}
Liang, D., Hure, J., Courcelle, A., Shawish, S.~E., Tanguy, B., 2021. A
  micromechanical analysis of intergranular stress corrosion cracking of an
  irradiated austenitic stainless steel. Acta Materialia 204, 116482.

\bibitem[{Luster and Morris(1995)}]{Luster1995}
Luster, J., Morris, M.~A., 1995. {Compatibility of deformation in two-phase
  Ti-Al alloys: Dependence on microstructure and orientation relationships}.
  Metallurgical and Materials Transactions A 26~(7), 1745--1756.

\bibitem[{McMurtrey et~al.(2015)McMurtrey, Cui, Robertson, Farkas, and
  Was}]{MCR15}
McMurtrey, M., Cui, B., Robertson, I., Farkas, D., Was, G., 2015. Mechanism of
  dislocation channel-induced irradiation assisted stress corrosion crack
  initiation in austenitic stainless steel. Current Opinion in Solid State and
  Materials Science 19, 305 -- 314.

\bibitem[{Narutani and Takamura(1991)}]{narutani1991grain}
Narutani, T., Takamura, J., 1991. {Grain-size strengthening in terms of
  dislocation density measured by resistivity}. Acta Metallurgica et Materialia
  39~(8), 2037--2049.

\bibitem[{Nye(1953)}]{Nye1953}
Nye, J.~F., 1953. {Some geometrical relations in dislocated crystals}. Acta
  Metallurgica 1~(2), 153--162.

\bibitem[{Petch(1953)}]{petch1953cleavage}
Petch, N.~J., 1953. {The cleavage strength of polycrystals, Journal of Iron and
  Steel Institute}. Journal of Iron and Steel Institute 174~(1), 25--28.

\bibitem[{Quey et~al.(2011)Quey, Dawson, and Barbe}]{Quey2011a}
Quey, R., Dawson, P.~R., Barbe, F., 2011. {Large-scale 3D random polycrystals
  for the finite element method: Generation, meshing and remeshing}. Computer
  Methods in Applied Mechanics and Engineering 200~(17-20), 1729--1745.

\bibitem[{Raj and Pharr(1986)}]{raj1986compilation}
Raj, S.~V., Pharr, G.~M., 1986. {A compilation and analysis of data for the
  stress dependence of the subgrain size}. Materials Science and Engineering
  81, 217--237.

\bibitem[{Read and Shockley(1950)}]{Read1950a}
Read, W.~T., Shockley, W., 1950. {Dislocation models of crystal grain
  boundaries}. Physical Review Letters 78, 275--289.

\bibitem[{Rubio et~al.(2019)Rubio, Haouala, and Llorca}]{Rubio2019}
Rubio, R.~A., Haouala, S., Llorca, J., 2019. {Grain boundary strengthening of
  FCC polycrystals}. Journal of Materials Research 34~(13), 2263--2274.

\bibitem[{Sauzay and Kubin(2011)}]{sauzay2011scaling}
Sauzay, M., Kubin, L.~P., 2011. {Scaling laws for dislocation microstructures
  in monotonic and cyclic deformation of fcc metals}. Progress in Materials
  Science 56~(6), 725--784.

\bibitem[{Schwartz et~al.(2009)Schwartz, Kumar, Adams, and Field}]{S2009}
Schwartz, A.~J., Kumar, M., Adams, B.~L., Field, D.~P., 2009. Electron
  backscatter diffraction in materials science. Vol.~2. Springer.

\bibitem[{Segurado et~al.(2018)Segurado, Lebensohn, and Llorca}]{Segurado2018}
Segurado, J., Lebensohn, R.~A., Llorca, J., 2018. {Computational Homogenization
  of Polycrystals}. Advances in Applied Mechanics 51, 1--114.

\bibitem[{Shen et~al.(1988)Shen, Wagoner, and Clark}]{Shen1988}
Shen, Z., Wagoner, R.~H., Clark, W.~A., 1988. {Dislocation and grain boundary
  interactions in metals}. Acta Metallurgica 36~(12), 3231--3242.

\bibitem[{Taylor(1934)}]{taylor1934mechanism}
Taylor, G.~I., 1934. {The mechanism of plastic deformation of crystals. Part
  II. Comparison with observations}. Proceedings of the Royal Society of London
  A 145~(855), 388--404.

\bibitem[{Wei and Anand(2004)}]{WA04}
Wei, Y.~J., Anand, L., 2004. Grain-boundary sliding and separation in
  polycrystalline metals: application to nanocrystalline fcc metals. Journal of
  the Mechanics and Physics of Solids 52, 2587 -- 2616.

\bibitem[{Zaiser and Sandfeld(2014)}]{zaiser2014scaling}
Zaiser, M., Sandfeld, S., 2014. {Scaling properties of dislocation simulations
  in the similitude regime}. Modelling and Simulation in Materials Science and
  Engineering 22~(6), 65012.

\bibitem[{Zhao et~al.(2020)Zhao, Bieler, LLorca, and Eisenlohr}]{ZBL20}
Zhao, Z., Bieler, T.~R., LLorca, J., Eisenlohr, P., 2020. Grain boundary slip
  transfer classification and metric selection with artificial neural networks.
  Scripta Materialia 185, 71 -- 75.

\end{thebibliography}

\end{document}